%% file: main.tex
\documentclass[twocolumn,showpacs,superscriptaddress,email,floatfix,longbibliography,prl]{revtex4-2}

\usepackage[english]{babel}
\usepackage{braket}

\usepackage{amsmath}
\usepackage{graphicx}
\usepackage[colorlinks=true, allcolors=blue]{hyperref}
\usepackage{comment}
\usepackage{subfiles} % Best loaded last in the preamble

\begin{document}
\title{Hybrid sub- and superradiant states in emitter arrays with quantized motion}
%\title{Sub- and superradiance in emitter arrays with quantized motion}
\author{Beatriz Olmos}
\affiliation{Institut f\"ur Theoretische Physik and Center for Integrated Quantum Science and Technology, Universit\"at Tübingen, Auf der Morgenstelle 14, 72076 T\"ubingen, Germany}
\author{Igor Lesanovsky}
\affiliation{Institut f\"ur Theoretische Physik and Center for Integrated Quantum Science and Technology, Universit\"at Tübingen, Auf der Morgenstelle 14, 72076 T\"ubingen, Germany}
\affiliation{School of Physics and Astronomy and Centre for the Mathematics and Theoretical Physics of Quantum Non-Equilibrium Systems, The University of Nottingham, Nottingham, NG7 2RD, United Kingdom}

\begin{abstract}
Ensembles of dipolar emitters which couple collectively to the radiation field display sub- and superradiance. These terms refer to a reduction or an enhancement of photon emission rates due to the interference of emission channels. Arrays of trapped neutral atoms constitute a promising platform for harnessing this phenomenon in technological applications, e.g. for excitation storage, single-photon switches and mirrors. However, vibrational motion of the atoms within their traps leads to position fluctuations that entangle the motion and the internal atomic degrees of freedom, which is expected to affect the collective photon emission. We develop here a theory for collective atom-light coupling in the presence of this quantized motion within the Lamb-Dicke limit. We show the existence of sub- and superradiant states, which are hybrids of electronic and vibrational excitations and explore their properties for analytically and numerically efficiently solvable cases.% and show that some of them these two degrees of freedom factorise, which establishes sub- and superradiance as robust quantum phenomena.
\end{abstract}

\maketitle

\textit{Introduction.} Atomic ensembles couple collectively to the radiation field, which leads to intricate open system dynamics. On the one hand, the exchange of virtual photons among atoms induces dipole-dipole interactions. On the other, the emission of photons stored within the ensemble takes place through collective emission channels that feature decay rates which are either larger (superradiant) or smaller (subradiant) than that of an isolated atom. These phenomena were theoretically predicted decades ago \cite{Dicke1954,Lehmberg1970,James1993}, and have been demonstrated experimentally in recent years not only in the case of atoms but also in molecules as well as quantum dots \cite{Bienaime2012,Pellegrino2014,Jenkins2016,Araujo2016,Guerin2016,Bromley2016,Ferioli2021,Lange2024,Tiranov2023,Ferioli2023}.

Collective effects become particularly pronounced when the atoms are periodically arranged, i.e. a lattice. Here, as the number of atoms is increased, the decay rates of the subradiant states can even approach zero due to the almost perfect interference achieved by the periodic configuration \cite{Asenjo2017,Sierra2022,Cech2023,Holzinger2024}. Proposals have been put forward to exploit the emergent long lifetimes for the realization of quantum storage and transport of photons \cite{Olmos2013,Jen2016,Facchinetti2016,Manzoni2018,Needham2019,Ballantine2020,Ballantine2021,Ballantine2022}, enhanced quantum metrology \cite{Ostermann2013,Reitz2022,Qu2019,Pineiro2020} or the realization of single-layer atomic mirrors \cite{Bettles2016,Rui2020,Buckley2022,Srakaew2023,Ruostekoski2023}. %However, for these proposals to be realised in experiment, the robustness of subradiance against disorder must be tested. 
For those and other applications it is important that subradiance is a robust, i.e. not fine-tuned, phenomenon. In particular, since this collective effect strongly depends on the geometric arrangement of particles, the presence of motion (which is coupled to the atomic internal degrees of freedom) may destroy the underlying interference mechanism. Often, the impact of the coupling between internal degrees of freedom and external motion is modelled via disorder, i.e., by  sampling the atomic positions from a Gaussian distribution %centered on each lattice site with standard deviation related to the lattice trap depth, 
and averaging over many realizations  \cite{Olmos2013,Bettles2016,Gjonbalaj2024}. However, a faithful description that considers the vibrational atomic motion as a genuine quantum degree of freedom becomes more and more important the better the experimental control of these (hybrid) quantum systems \cite{Palmer2010,Damanet2016,Guimond2019,Rusconi2021,Rubiesbigorda2024}.

\begin{figure}[t]
\centering
\includegraphics[width=\columnwidth]{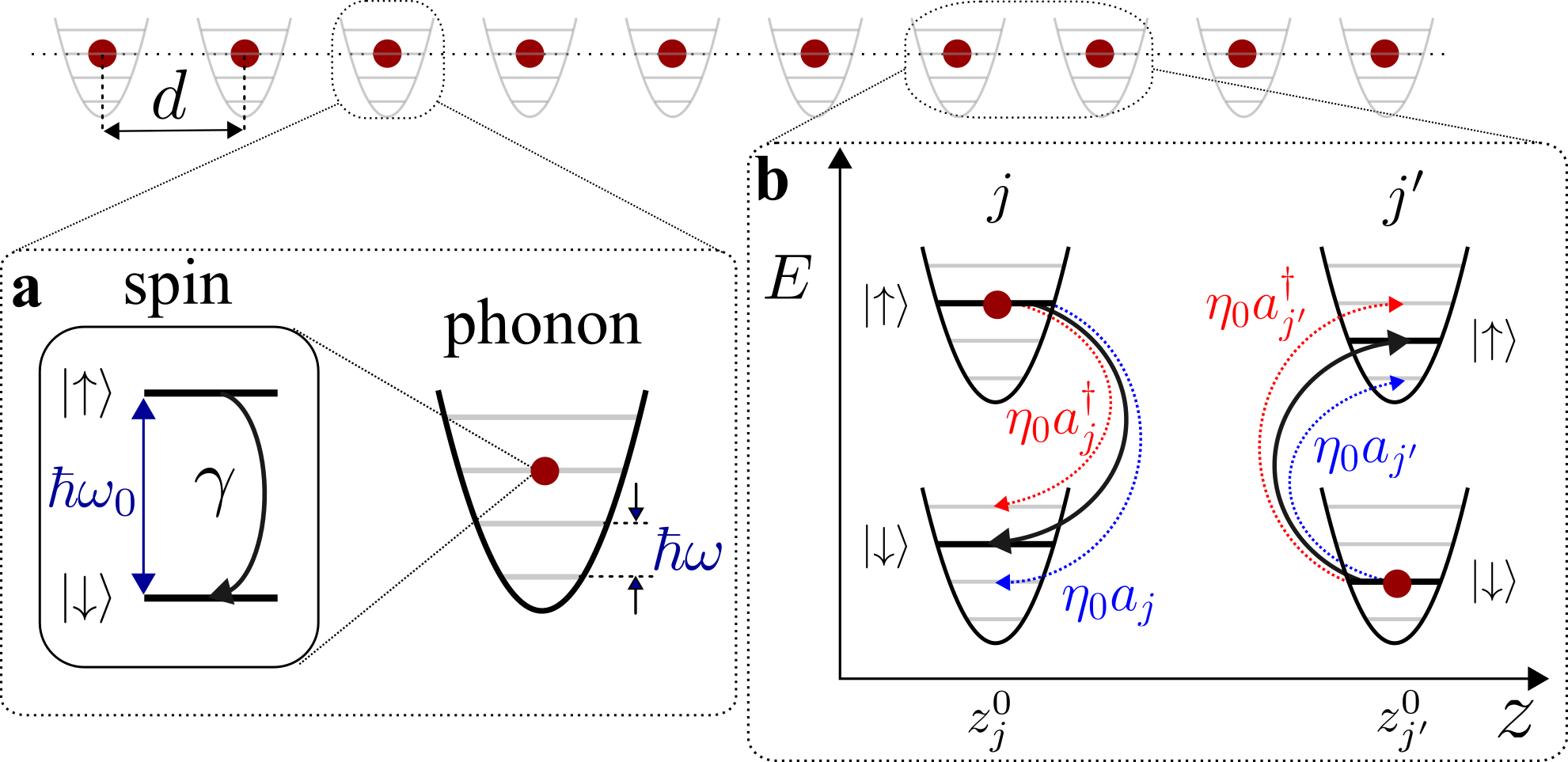}
\caption{\label{fig:fig1} \textit{Spin-phonon coupling.} \textbf{a:} The internal degrees of freedom of a chain of atoms with two electronic levels, separated by an energy $\hbar\omega_0$, are coupled to the radiation field and to the quantized motional degrees of freedom (one-dimensional vibrations in a trap, represented by phonons of energy $\hbar\omega$). \textbf{b:} Collective photon emission and dipole-dipole exchange occurs between two emitters at positions $z_j^0$ and $z_{j'}^0$ (black arrows). Photon recoil leads to processes that can change the vibrational state by creating or annihilating one phonon (red and blue arrows, respectively). The rates of these processes are proportional to the Lamb-Dicke parameter $\eta_0$.}
\end{figure}

In this paper, we develop a theory of hybrid sub- and superradiant states for atomic ensembles with coupled internal electronic degrees of freedom (henceforth referred to as spin) and external motion (phonons). We derive a Lindblad master equation within the Lamb-Dicke limit from first principles and analytically solve, first, the particular case of two atoms. Here, we show that for a certain atomic separation and/or choice of the transition dipole moment orientation, the sub- and superradiant decay rates as well as the dipole-dipole interactions are not affected by the spin-phonon coupling. Considering an atomic chain, we find that the latter coupling generally hybridizes the two sets of degrees of freedom. Nevertheless, also in this hybrid system a class of states can be identified, which are separable, i.e., where the spin and phonon states factorize. Our study thus sheds light on the various manifestations of sub- and superradiant states in a composite quantum system, which models many experimentally relevant scenarios.

\textit{Model and equations of motion.} We consider a system of $N$ two-level emitters (here neutral atoms) in a one-dimensional configuration coupled to the free radiation field. The internal degrees of freedom of each atom can be considered as those of a fictitious spin-$1/2$ particle, with lower ($\ket{\downarrow}$) and upper ($\ket{\uparrow}$) states separated by an energy $\hbar\omega_0$, as shown in Fig. \ref{fig:fig1}a. Each of these atoms is assumed to be at the position $\mathbf{r}_j=z_j\hat{z}$, trapped in a harmonic potential centered at $\mathbf{r}_j^0=z_j^0\hat{z}$. The quantized motion within each trap is strictly assumed to take place in the direction of the chain, i.e. the $z$-axis. For the $j$-th atom, this motion is represented via the bosonic phonon operators $a_j$ and $a_j^\dagger$, which annihilate and create vibrational excitations of energy $\hbar\omega\ll\hbar\omega_0$, respectively (see Fig. \ref{fig:fig1}a). The displacement of the position of the atom with respect to the center of its trap can be written in terms of these operators, such that $z_j=z_j^0+z_\mathrm{ho}(a_j+a^\dag_j)/\sqrt{2}$, where $z_\mathrm{ho}=\sqrt{\hbar/m\omega}$ is the harmonic oscillator length and $m$ the atomic mass.

Within the dipole approximation, the Hamiltonian that describes this system is given by \cite{Lehmberg1970}
\begin{equation*}
    \begin{split}
H=&\hbar\sum_{\mathbf{k},\lambda} \nu_kc^\dag_{\mathbf{k}\lambda} c_{\mathbf{k}\lambda}+\hbar\omega_0\sum_{j=1}^N\sigma_j^\dag\sigma_j+\hbar\omega\sum_{j=1}^N a_j^\dag a_j\\
&-\hbar\sum_{j,\mathbf{k},\lambda}g_\mathbf{k}^\lambda(\sigma_j^\dag+\sigma_j)(c_{\mathbf{k}\lambda}e^{\mathrm{i}\mathbf{k}\cdot\mathbf{r}_j}+c^\dag_{\mathbf{k}\lambda}e^{-\mathrm{i}\mathbf{k}\cdot\mathbf{r}_j}).
\end{split}
\end{equation*}
Here, we have introduced the the spin ladder operators $\sigma_j=\ket{\downarrow}_j\!\bra{\uparrow}$ and $\sigma_j^\dag=\ket{\uparrow}_j\!\bra{\downarrow}$, the bosonic creation and annihilation operators $c_{\mathbf{k}\lambda}^\dag$ and $c_{\mathbf{k}\lambda}$ of a photon with momentum $\mathbf{k}$, energy $\hbar\nu_k=\hbar c|\mathbf{k}|$ and polarization $\lambda=1,2$. Moreover,the coupling constant between the emitters and the radiation field  is given by $g^\lambda_\mathbf{k}=\mathbf{d}\cdot\hat{\epsilon}^\lambda_\mathbf{k}\sqrt{\frac{\nu_k}{2\varepsilon_0\hbar V}}$, with $\mathbf{d}$, $\hat{\epsilon}^\lambda_\mathbf{k}$, and $V$ being the atomic transition dipole moment, the field polarization unit vector and the quantization volume, respectively.

%To parametrize the phononic coupling, we introduce the parameter $\eta=\mathbf{k}\cdot\hat{z}z_\mathrm{ho}/\sqrt{2}\ll1$. In the interaction picture with respect to the atomic, field and phonon frequencies, and expanding around $\eta=0$, the final Hamiltonian can be written up to second order in $\eta$ as $H\approx H_0+H_1+H_2$ where
%\begin{align*}
%    H_0=&-\hbar\sum_{j,\mathbf{k}}g_\mathbf{k}A_j(t)\left[B_{\mathbf{k}j}(t)+B^\dag_{\mathbf{k}j}(t)\right]\\
%    H_1=&-\mathrm{i}\hbar\sum_{j,\mathbf{k}}\eta g_\mathbf{k}A'_j(t)\left[B_{\mathbf{k}j}(t)-B^\dag_{\mathbf{k}j}(t)\right]\\
%    H_2=&\hbar\sum_{j,\mathbf{k}}\frac{\eta^2}{2}g_\mathbf{k}A''_j(t)\left[B_{\mathbf{k}j}(t)+B^\dag_{\mathbf{k}j}(t)\right]
%\end{align*}
%where we have defined $A_j(t)=e^{i\omega_0 t}\sigma_j^\dag+\mathrm{h.c.}$, $A'_j(t)=A_j(t)(e^{\mathrm{i}\omega t}a_j^\dag+\mathrm{h.c.})$, $A''_j(t)=A_j(t)\left[a_j^\dag a_j+e^{2\mathrm{i}\omega t}(a_j^\dag)^2+\mathrm{h.c.}\right]$ and $B_{\mathbf{k}j}(t)=p_\mathbf{k}e^{\mathrm{i}(\mathbf{k}\cdot\mathbf{r}_j^0-\nu_kt)}$.

%\textbf{[Reads weird. Shouldn't this sentence come later (and be split into two sentences)? I would start with the next sentence. I would also explicitly say that details are found in the SM:]}  

\begin{figure}[t!]
\centering
\includegraphics[width=\columnwidth]{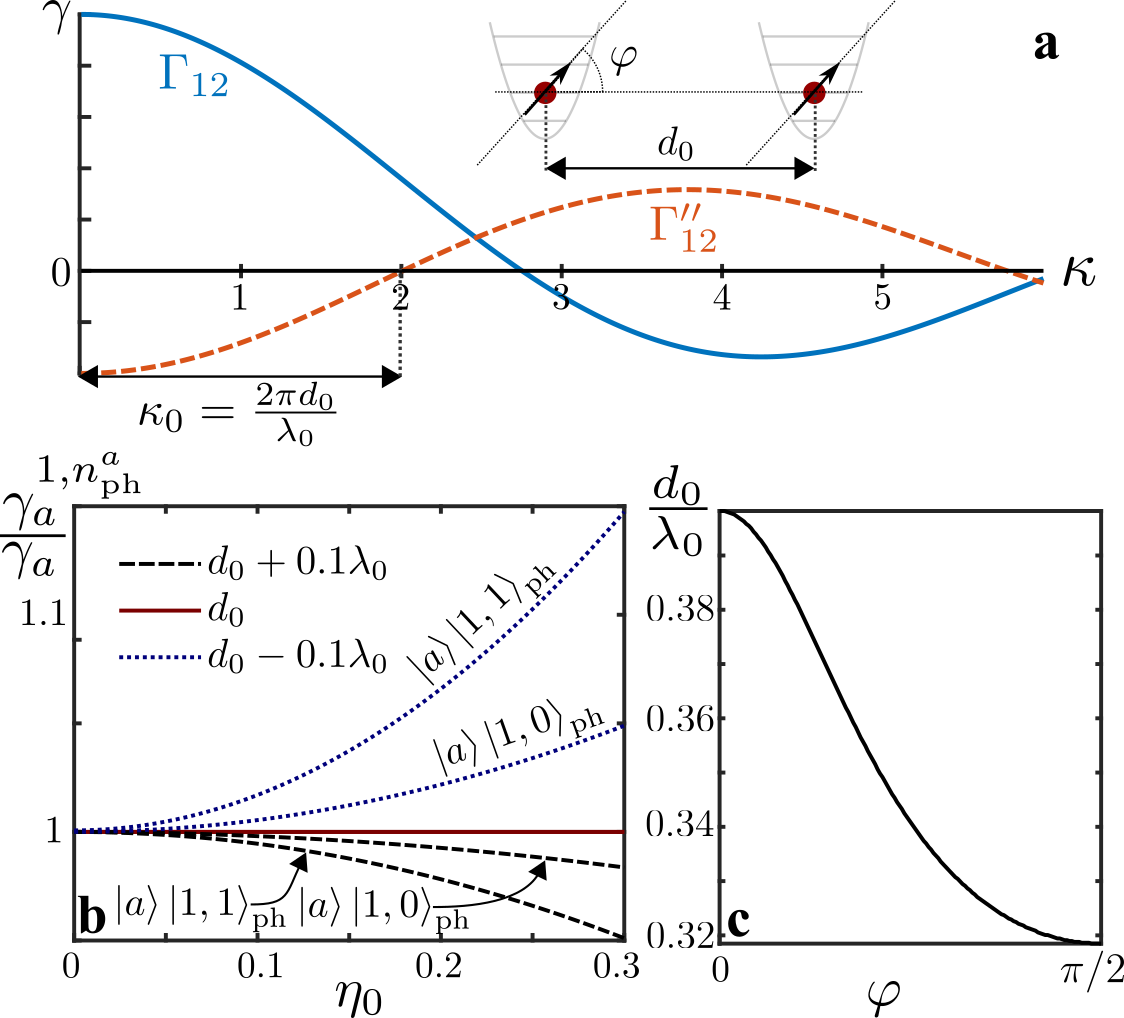}
\caption{\label{fig:fig2} \textit{Two atoms.} \textbf{a:} For two atoms, the change of the collective decay rates due to the spin-motion coupling is proportional to the second derivative of the function $\Gamma_{12}$ with respect to $\kappa\equiv2\pi d/\lambda_0$. This function has a zero at $\kappa=\kappa_0$ (and distance $d=d_0$). \textbf{b:} Collective decay rates $\gamma_a^{1,n_\mathrm{ph}^a}$ of the product states $\ket{a}\ket{1,n_\mathrm{ph}^a}$, where the atoms are in the antisymmetric state $\ket{a}=(\ket{\uparrow \downarrow}-\ket{\downarrow\uparrow})/\sqrt{2}$ and the single phonon in the state $\ket{1,0}_\mathrm{ph}=a_s^\dag\ket{0}_\mathrm{ph}$ or $\ket{1,1}_\mathrm{ph}=a_a^\dag\ket{0}_\mathrm{ph}$, as a function of $\eta_0$. The rates are divided by their value at $\eta_0=0$, $\gamma_a$. When setting the interatomic distance to $d=d_0$, the rates are independent of $\eta_0$. \textbf{c:} Dependence of the interatomic distance at which the second derivative $\Gamma''_{12}$ vanishes, $d_0$, on the angle $\varphi$ between the dipole moments (black arrows in the inset in \textbf{a}) and the axis connecting the atoms.}
\end{figure}

By eliminating the radiation field, and under the Born and Markov approximations \cite{Lehmberg1970,Manzano2020}, we obtain an equation of motion for the dynamics of the spin and phonon degrees of freedom only (see \cite{SM} for details). We consider the situation where the atoms are tightly trapped, such that $z_\mathrm{ho}\ll\lambda_0$, where $\lambda_0=2\pi c/\omega_0$ is the wavelength of the atomic transition (Lamb-Dicke parameter regime). This allows us to expand in the small Lamb-Dicke parameter $\eta_0=k_0z_\mathrm{ho}/\sqrt{2}\ll1$ with $k_0=2\pi/\lambda_0$, up to order $\eta_0^2$. The dynamics is then governed by the Lindblad master equation
\begin{equation}\label{eq:ME}
\begin{split}
    \dot{\rho}=&\sum_{m,m'=1}^{4N}\tilde{\Gamma}_{mm'}\left(J_{m'}\rho J_{m}^\dag-\frac{1}{2}\left\{J_m^\dag J_{m'},\rho\right\} \right)\\&+\mathrm{i}\sum_{m\neq m'}\tilde{V}_{mm'}\left[J_{m}^\dag J_{m'},\rho\right].
\end{split}
\end{equation}
This equation accounts for collective photon emission and dipole-dipole exchange interactions (first and second term, respectively). The two matrices $\tilde{\Gamma}$ and $\tilde{V}$ are $4N\times4N$ block matrices of the form
\begin{equation*}
    \tilde{\Gamma}=\left(\begin{array}{cccc}
        \Gamma & 0 & 0 & \frac{\eta_0^2}{2} \Gamma'' \\
         0 & -\eta_0^2 \Gamma'' & 0 & 0 \\ 
         0 & 0 & -\eta_0^2\Gamma'' & 0 \\
         \frac{\eta_0^2}{2}\Gamma'' & 0 & 0 & 0 
    \end{array}\right).
\end{equation*}
Here, the $N\times N$ block matrix $\Gamma$ has the entries
\begin{equation*}
    \Gamma_{jj'}=\frac{3\gamma}{2}\left[f_\varphi\frac{\sin{\kappa_{jj'}}}{\kappa_{jj'}}+g_\varphi\left(\frac{\cos{\kappa_{jj'}}}{\kappa_{jj'}^2}-\frac{\sin{\kappa_{jj'}}}{\kappa_{jj'}^3}\right)\right],
\end{equation*}
where $\gamma$ is the single-atom decay rate, $\kappa_{jj'}=k_0z^0_{jj'}\equiv k_0|z^0_{j}-z^0_{j'}|$ the reduced distance between two atoms, and $\Gamma''_{jj'}=\partial^2_{\kappa_{jj'}}\Gamma_{jj'}$. Moreover, here $f_\varphi=\sin^2{\varphi}$, $g_\varphi=1-3\cos^2{\varphi}$, where $\cos{\varphi}=\hat{\mathbf{d}}\cdot\hat{r}^0_{jj'}$ is the relative angle between the transition dipole moments and the line connecting the two atoms. We assume this angle to be the same throughout the chain (see Fig. \ref{fig:fig2}a). The same structure applies to the dipole-dipole interaction matrix $\tilde{V}$, whose entries are
\begin{equation*}
    V_{jj'}=\frac{3\gamma}{4}\left[f_\varphi\frac{\cos{\kappa_{jj'}}}{\kappa_{jj'}}-g_\varphi\left(\frac{\sin{\kappa_{jj'}}}{\kappa_{jj'}^2}+\frac{\cos{\kappa_{jj'}}}{\kappa_{jj'}^3}\right)\right],
\end{equation*}
for $j\neq j'$ and $V_{jj}=0$, and $V''_{jj'}=\partial^2_{\kappa_{jj'}} V_{jj'}$. The master equation (\ref{eq:ME}) contains four sets of jump operators:
\begin{equation*}
    J_m=\begin{cases}
    \sigma_j & 1\leq m\leq N\\
    \sigma_{j} a_{j} & N+1\leq m\leq 2N\\
    \sigma_{j} a^\dagger_{j} & 2N+1\leq m\leq 3N\\
    \sigma_{j}(1+2a_{j}^\dag a_{j}) & 3N+1\leq m\leq 4N,
    \end{cases}
\end{equation*}
with indices $j=1\dots N$. The first $N$ terms ($m,m'=1,\dots,N$) of the master equation \eqref{eq:ME} recover the well-established equation for two-level atoms coupled to the radiation field in the absence of spin-phonon coupling \cite{Lehmberg1970,James1993,Olmos2013,Asenjo2017,Needham2019}. The rest of the equation yields the leading correction (of order $\eta_0^2$) to the dynamics due to the recoil of the emitters under photon emission. The processes that accompany these corrections correspond either to the loss or the gain of a phonon in each spin transition (see Fig. \ref{fig:fig1}b). A further effect of the same order is the renormalization of the spin transition rates, which acquire a dependence on the number of phonons.

In the following we focus on the dynamics when at most one spin excitation, $\ket{\uparrow}$, is present in the whole system, also known as the linear optics regime \cite{Bettles2016,Needham2019,Ballantine2020}. Here, the spin ladder operators $\sigma_j$ and $\sigma_j^\dagger$ can be substituted by the bosonic annihilation and creation operators $b_j$ and $b_j^\dag$, respectively. In this regime the dynamics is governed by the non-hermitian Hamiltonian $H^\mathrm{eff}=\sum_{jj'}H^\text{eff}_{jj'}$ with
\begin{equation}
\begin{split}
    H^\text{eff}_{jj'}=& b_{j}^\dag b_{j'}\left[M_{jj'}+\eta_0^2M''_{jj'}\left(1-\delta_{jj'}\right.\right.\\
    &\left.\left.+a_{j'}^\dag a_{j'}+a_{j}^\dag a_{j}-a_{j}^\dag a_{j'}-a_{j'}^\dag a_{j}\right)\right]
\end{split}\label{eq:Heff}
\end{equation}
where $M_{jj'}=V_{jj'}-\mathrm{i} \Gamma_{jj'}/2$, and $M''_{jj'}=\partial^2_{\kappa_{jj'}} M_{jj'}$. The real and imaginary parts of the eigenvalues of $H^\mathrm{eff}$ represent the collective energy shifts and decay rates, respectively. Note, that this Hamiltonian conserves the total number of phonons, $n_\mathrm{ph}$, that are contained in the lattice.

%, which we will consider in the following.

\textit{Sub- and superradiant states of two atoms.} Let us first analyze the case of two atoms, which are separated by a distance $d$. Not only can this system be solved fully analytically, but one can also decouple spin and motion. Let us rewrite the effective Hamiltonian \eqref{eq:Heff} for $N=2$ in diagonal form. To do so, we introduce the symmetric and antisymmetric combinations of bosonic operators representing the atomic excitation, $b_{s/a}=(b_1\pm b_2)/\sqrt{2}$, and the phonons, $a_{s/a}=(a_1\pm a_2)/\sqrt{2}$. This yields
% \begin{equation*}
%     H^\text{eff}=-\frac{\mathrm{i\gamma}}{2}+(b_s^\dag b_s-b_a^\dag b_a)\left[M_{12}+\eta_0^2M''_{12}\left(1+2a_a^\dag a_a\right)\right].
% \end{equation*}
\begin{equation}\label{eq:HeffN2}
    \begin{split}
    H^\text{eff}=&M_{11}+(b_s^\dag b_s-b_a^\dag b_a)\left(M_{12}
    +\eta_0^2M''_{12}\right)\\
    &+2\eta_0^2(b_s^\dag b_s-b_a^\dag b_a)M''_{12}a_a^\dag a_a,
    \end{split}
\end{equation}
where $M_{11}=M_{22}=-\mathrm{i}\gamma$/2. This expression reveals that all eigenstates of the Hamiltonian are separable (product states) in the atomic and phonon degrees of freedom. The atomic eigenstates are the symmetric and antisymmetric superpositions  $\ket{s/a}=b_{s/a}^\dag\ket{\downarrow\downarrow}=(\ket{\uparrow \downarrow}\pm\ket{\downarrow\uparrow})/\sqrt{2}$. The phonon eigenstates are 
$\ket{n_\mathrm{ph},n_\mathrm{ph}^a}={a_s^\dag}^{(n_\mathrm{ph}-n_\mathrm{ph}^a)}{a_a^\dag}^{n_\mathrm{ph}^a}\ket{0}_\mathrm{ph}$, where $n_\mathrm{ph}$ is the total number of phonons and $n_\mathrm{ph}^a$ is the number of phonons in the antisymmetric mode. The corresponding collective decay rates are given by $\gamma_{s/a}^{n_\mathrm{ph},n_\mathrm{ph}^a}=\gamma\pm\Gamma_{12}\pm\eta_0^2(2n_\mathrm{ph}^a+1)\Gamma''_{12}$.
When $\eta_0=0$ there exist only two degenerate rates, namely $\gamma_{s/a}^{n_\mathrm{ph},n_\mathrm{ph}^a}\equiv\gamma_{s/a}=\gamma\pm\Gamma_{12}$, governing superradiant ($\gamma_s>\gamma$) and subradiant ($\gamma_a<\gamma$) emission. For $\eta_0\neq 0$, we find that all rates change by an amount proportional to $\Gamma''_{12}$. %The rate also depends on the phonon occupation of the antisymmetric mode.
Note that, unlike in long-range interacting systems, where the interactions decay monotonically with the distance, the behaviour of both $\Gamma_{12}$ and $V_{12}$ with the reduced distance $\kappa$ is non-monotonic (see Fig. \ref{fig:fig2}a). Hence, here it is possible to find an interatomic distance $d_0$ where the second derivative $\Gamma''_{12}$ vanishes. Here, all decay rates remain unaffected by the spin-motion coupling independently of both the phonon number and the Lamb-Dicke parameter $\eta_0$. %This means that here not only are all eigenstates separable in spin and phonon degrees of freedom, \textbf{[Do we need the second half of this sentence?:]} but also both sub- and superradiant states are virtually impervious to motional effects (up to second order in $\eta_0$). 
%\textbf{[This is a bit unclear: e.g. which phonons are you talking about? What does "two spin antisymmetric collective decay rates" mean?:]} 
To illustrate this, we show in Fig. \ref{fig:fig2}b the collective decay rates of two atoms for three values of the distance $d$ and $n_\mathrm{ph}=1$. Indeed, when $d=d_0$, the rates are independent of $\eta_0$.

%In particular, we plot the decay rates $\gamma_a^{s/a}$ of the two subradiant eigenstates $\ket{a}a^\dagger_{s/a}\ket{0}_\mathrm{ph}$ with $\ket{a}=b_a^\dag\ket{\downarrow\downarrow}$, for three values of the distance $d$. Indeed, when $d=d_0$, the two rates remain unchanged as $\eta_0$ increases.

Note, moreover, that, as shown in Fig. \ref{fig:fig2}c, the distance $d_0$ at which the second derivative $\Gamma''_{12}$ vanishes depends as well on the relative angle $\varphi$ between the transition dipole moments and the axis connecting the two atoms. Interestingly, in the case of $\varphi=\pi/2$ (dipoles perpendicular to the $z$-axis), and when choosing $\kappa_0=2\pi d_0/\lambda_0=2$, not only $\Gamma_{12}$ but also $V_{12}$ possesses an inflection point. Hence, in this special case not only the decay rates but also the energy shifts of the collective eigenstates are left unaffected by the spin-motion coupling.

\textit{Many atoms.} When $N>2$, the problem is in general no longer analytically solvable. However, we can gain an intuition of the physics by considering an infinite chain of atoms, i.e. $N$ tending to infinity (see e.g. Refs. \cite{Asenjo2017,Sierra2022,Cech2023,Holzinger2024}). In this limit, one can diagonalize the effective Hamiltonian by introducing the Fourier transformed operators
\begin{equation*}
    b_j=\frac{1}{\sqrt{N}}\sum_{q=-\pi/d}^{\pi/d}e^{\mathrm{i}z_j^0q}\tilde{b}_q,\quad
    a_j=\frac{1}{\sqrt{N}}\sum_{p=-\pi/d}^{\pi/d}e^{\mathrm{i}z_j^0p}\tilde{a}_p
\end{equation*}
with $z_j^0=jd$ and $j=-N/2,\dots,N/2$.
The effective Hamiltonian \eqref{eq:Heff} becomes
\begin{equation}
\begin{split}\label{eq:HeffFourier}
    H^\text{eff}=&\sum_q \tilde{b}^\dag_q \tilde{b}_q (\tilde{M}_q+\eta_0^2\tilde{M}''_q) -\eta_0^2M''_{11}\\
    &+\frac{\eta_0^2}{N}\!\sum_{q,p,p'}\tilde{b}^\dag_{q+p'}\tilde{b}_{q+p}\tilde{a}^\dag_{p'}\tilde{a}_p\\
    &\times\left(\tilde{M}''_{q+p}+\tilde{M}''_{q+p'}-\tilde{M}''_q-\tilde{M}''_{q-p+p'}\right),
    \end{split}
\end{equation}
where we have used that $M''_{jj}\equiv M''_{11}$ for all $j$. Here, we have also introduced the Fourier transform of the matrix $M$ as $\tilde{M}_q=\sum_{jj'} e^{\mathrm{i}q(z^0_{j}-z^0_{j'})}M_{jj'}$, where the sum runs over $z^0_{j}-z^0_{j'}=-Nd/2,\dots,Nd/2$, and equivalently for the second derivative, $\tilde{M}''_q$.

\begin{figure}[t]
\includegraphics[width=\columnwidth]{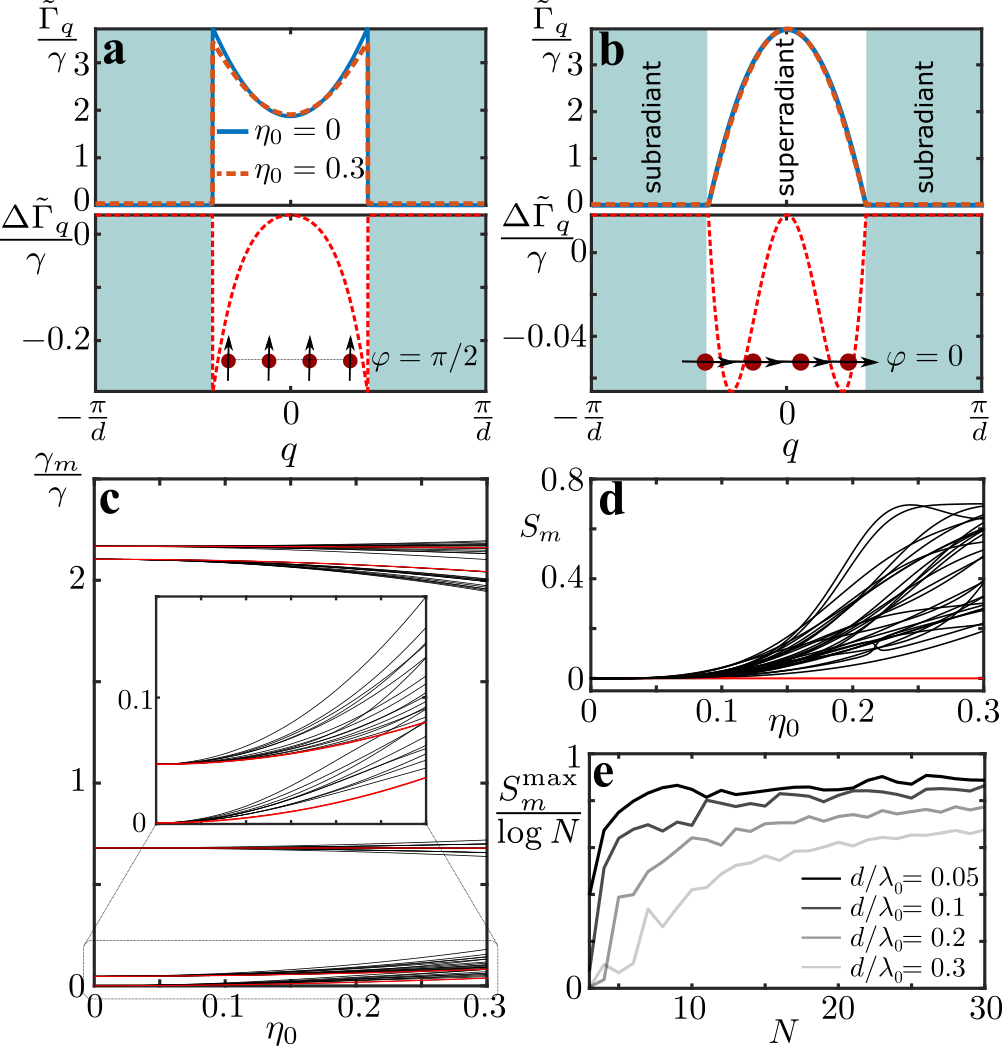}
\caption{\label{fig:fig3} \textit{Atomic chain}. \textbf{a} and \textbf{b}: The upper panels show the decay rates of an infinite chain with $d/\lambda_0=0.2$, for $\eta_0=0$, i.e., $\tilde{\Gamma}_q=-2\mathrm{Im}(\tilde{M}_q)$, and for the separable states $\ket{q}(a^\dag_0)^{n_\mathrm{ph}}\ket{0}_\mathrm{ph}$ at $\eta_0=0.3$, i.e. $\tilde{\Gamma}_q=-2\mathrm{Im}(\tilde{E}_q)$. In the lower panels we show the difference of the decay rates with and without spin-phonon coupling, $\Delta\tilde{\Gamma}_q=\tilde{\Gamma}_q(\eta_0=0.3)-\tilde{\Gamma}_q(\eta_0=0)$. The blue shaded areas cover the quasimomenta $|q|>2\pi/\lambda_0$, where all eigenstates are subradiant. \textbf{c:} Decay rates for a chain of $N=5$ atoms, containing $n_\mathrm{ph}=2$ phonons. The red lines indicate the product states. \textbf{d:} Entanglement entropy for the states shown in \textbf{c}. $S_m=0$ for all values of $\eta_0$ for the separable states (red lines). \textbf{e:} Maximum entanglement entropy at $\eta_0=0.3$ in a chain with $n_\mathrm{ph}=1$ phonon, as a function of the system size $N$. As the nearest neighbor distance $d/\lambda_0$ is reduced, the entropy approaches faster its maximum possible value, $\log{N}$.}
\end{figure}

For $\eta_0=0$, the quasi-momentum $q$ is a good quantum number and all eigenstates of the effective Hamiltonian \eqref{eq:HeffFourier} are products of phonon states and the atomic (single) excitation states $\ket{q}=\tilde{b}_q^\dag\ket{\downarrow,\dots,\downarrow}$, with degenerate decay rates $\tilde{\Gamma}_q=-2\text{Im}(\tilde{M}_q)$. Here, when $d<\lambda_0/2$, it is well known that all eigenstates with $|q|>2\pi/\lambda_0$ are completely subradiant, i.e. $\tilde{\Gamma}_{|q|>2\pi/\lambda_0}=0$, while the rest, $|q|<2\pi/\lambda_0$, have a finite decay rate (see Figs. \ref{fig:fig3}a and b).
%In Figs. \ref{fig:fig3}a and b we show the decay rates $\tilde{\Gamma}_q=-2\text{Im}(\tilde{M}_q)$ of an infinite chain with $d/\lambda=0.2$, i.e. , for two values of the dipole angle $\varphi$.
As we increase $\eta_0$, $q$ is in general no longer a good quantum number. However, upon inspection of the effective Hamiltonian \eqref{eq:HeffFourier}, one can see that all product states of the form $\ket{q}(\tilde{a}^\dag_0)^{n_\mathrm{ph}}\ket{0}_\mathrm{ph}$ are eigenstates of the system. Here, $\tilde{a}^\dag_0=\sum_j a_j^\dag/\sqrt{N}$ is the creation operator of the center of mass motion, where the relative distance between the emitters is unchanged. The eigenvalues for these product states are given by
\begin{align*}
    \tilde{E}_q=\tilde{M}_q+\eta_0^2(\tilde{M}''_q-M''_{11}).
\end{align*}
In Figs. \ref{fig:fig3}a and b we contrast the collective decay rates in the cases with and without spin-phonon coupling.
%The decay rates $\tilde{\Gamma}_q=-2\text{Im}(\tilde{E}_q)$ for these separable states \textbf{[Unnecessary detail, which makes the sentence hard to read. Just say that in the figure we contrast the cases with and without spin-phonon coupling.:]} for $\eta_0=0.3$ are compared to the $\eta_0=0$ case in Figs. \ref{fig:fig3}a and b an infinite chain with $d/\lambda=0.2$ and two values of the dipole angle $\varphi$.
Here we can see that, even though these eigenstates are separable into atomic and phonon degrees of freedom, the coupling still leads to a  renormalization of the decay rates proportional to $\eta_0^2$.

In general, however, the eigenstates of the effective Hamiltonian are nonseparable hybrids of atomic and vibrational excitations. In the following we investigate this hybridzation in a finite chain with open boundaries. In particular, we evaluate numerically, for a fixed number of atoms $N$ and phonons $n_\mathrm{ph}$, the eigenstates $\ket{\psi_m}$ and eigenvalues $E_m$ of the effective Hamiltonian \eqref{eq:Heff}, where $m= 1,2,\dots, {N+n_\mathrm{ph}-1 \choose n_\mathrm{ph}}$ with the binomial coefficient giving the Hilbert space dimension. For each state we then calculate the collective decay rate $\gamma_m=-2\mathrm{Im}(E_m)$ and the spin-motion entanglement. To quantify the latter, we trace out the phonon degrees of freedom of each state $\rho^m_\mathrm{tot}=\ket{\psi_m}\bra{\psi_m}$ and obtain the reduced (generally mixed) density matrix $\rho_m=\mathrm{Tr}_\mathrm{ph}(\rho^m_\mathrm{tot})$. Subsequently, we calculate the entanglement entropy of the hybrid (bipartite) atom-phonon system,
%with $m= 1\dots {N+n_\mathrm{ph}-1 \choose n_\mathrm{ph}}$, which we depict in Figure \ref{fig:fig3}c for $N=5$ atoms with $n_\mathrm{ph}=2$ phonons. Moreover, we are interested in the spin-motion entanglement in each eigenstate $\ket{\psi_m}$ of the effective Hamiltonian.
%As we will see in the following, the \textbf{[Which result exactly are you referring to?]} previous result is not an artifact of the translationally invariant situation, but also survives in small systems with open boundaries.
%In Figure \ref{fig:fig3}c we evaluate numerically the decay rates for a system of $N=5$ atoms with $n_\mathrm{ph}=2$ phonons. Here, we can observe, first, that again the perturbation of the decay rates as we increase $\eta_0$ is \textbf{[I don't like the term. Before you even used "extremely small". But this is expected, since $\eta$ is supposed to be small.:]} relatively small.
%To quantify the spin-motion entanglement in each eigenstate $\ket{\psi_m}$ of the effective Hamiltonian $H^\mathrm{eff}$, we trace out the phonon degrees of freedom  and obtain the reduced density matrix $\rho_m=\mathrm{Tr}_\mathrm{ph}(\rho^m_\mathrm{tot})$, where $\rho^m_\mathrm{tot}=\ket{\psi_m}\bra{\psi_m}$. While $\rho^m_\mathrm{tot}$ is a pure state, $\rho_m$ can be mixed if the spin and phonon degrees of freedom hybridize. We calculate for each eigenstate the entanglement entropy of the hybrid (bipartite) atom-phonon system,
\begin{equation*}
    S_m=-\rho_m\log\rho_m,
\end{equation*}
which is zero for any product state. In Figs. \ref{fig:fig3}c and d we show $\gamma_m$ and $S_m$ for an atomic chain with $N=5$ atoms with $n_\mathrm{ph}=2$ phonons. The majority of the eigenstates feature entanglement, i.e. $S_m\neq 0$, when $\eta_0$. Moreover, there exist $N$ eigenstates that are not entangled ($S_m=0$). These correspond to center of mass vibrations, similar to what was found in the infinite chain. Interestingly, when increasing the system size $N$, some of the hybridized states come close to saturating the entanglement entropy, i.e. they reach an entanglement entropy near the maximum allowed value, $\log{N}$. As depicted in Fig. \ref{fig:fig3}e, this saturation with increasing $N$ is faster the smaller the ratio $d/\lambda_0$ between the lattice constant and the wavelength of the atomic transition, where the interactions between the atoms become increasingly stronger. Nevertheless, also these heavily hybridized states clearly manifest super- and subradiance.

% Note, however, that even in this regime there always exist states which are separable and hence minimally unperturbed by this coupling.

%\textbf{[Convoluted sentence? I am not sure that $N=5$ is important here. In general, this reads a bit disordered. What do you want to say, what do we learn? Also, when referring to the figures, too much detail is given, I would say.:]} 

%As we can see in Figs. \ref{fig:fig3}c,d for $N=5$, as predicted in the translationally invariant case, in each of the $N$ manifolds of eigenstates for $\eta_0=0$ one state remains separable (the one that contains here two center of mass phonons, depicted in red). Moreover, the decay rates of these separable states are the least affected by the perturbation.
%The maximum value that the entanglement entropy can achieve is $\log{N}$, since $N$ is the dimension of the spin effective Hamiltonian. In Fig. \ref{fig:fig3}e we have calculated, for a system of $N$ atoms and $n_\mathrm{ph}=1$ phonon, the maximum value of $S_m$ at $\eta_0=0.3$. We observe that, as we increase $N$, the entropy eventually reaches its maximum possible value.

\textit{Conclusions and outlook.} We have investigated the impact of quantized motion on the collective decay of a chain of atoms coupled to the radiation field. Our findings show that super- and subradiance  are robust quantum phenomena, which also manifest in hybrid systems. In the future it would be interesting to use the derived master equation to go beyond the regime of a single atomic excitation studied here, and to investigate the impact of the spin-phonon coupling in a many-body setting (e.g. superradiance from a fully inverted state \cite{Sierra2022,Masson2022} or a strongly driven system \cite{Ott2013,Olmos2014}). Other future directions include the consideration of lattices with higher dimensions and with varying geometry, and to study their interplay with angular-dependent dipolar interactions that necessitate to go beyond the two-level approximation. Moreover, exploring the regime of strong coupling, i.e. beyond the Lamb-Dicke limit, appears to be a fruitful avenue for future studies.
%going beyond the averaging of disordered configurations, by considering the motion as a fully quantum degree of freedom. We have derived from first principles a Lindblad equation in the Lamb-Dicke regime that faithfully describes the dynamics of both atomic and motional degrees of freedom. Applied to a chain of atoms, we find sub and superradiant states in this fully quantised system, some of them factorising. Going beyond the single excitation regime to study the impact of the spin-phonon coupling in a many-body setting (e.g. superradiance from a fully inverted system) by means of our master equation \eqref{eq:ME} will be the obvious next step. 

\acknowledgments
\textit{Acknowledgments.}
We acknowledge funding from the Deutsche Forschungsgemeinschaft within the Grant No. 452935230 and the research units FOR5413 (Grant No. 465199066) and FOR5522 (Grant No. 499180199). This work was also supported by the QuantERA II programme (project CoQuaDis, DFG Grant No. 532763411) that has received funding from the EU H2020 research and innovation programme under GA No. 101017733.

%\bibliographystyle{alpha}
%\bibliography{sample}

%%%%%%%%%%%%%%%%%%%%%%%%%%%%%%%%%%%%%%%%%%%%%%%%%%%%%%%%%%%%%%%
%% COMMENT THIS OUT FOR PHYS REV VERSION
%%
%% YOU NEED IT IF YOU WOULD LIKE TO COMPILE
%% THE APPENDIX TO THE SAME PDF FOR ARXIV VERSION

\onecolumngrid
\clearpage
\subfile{supp.tex}
%%%%%%%%%%%%%%%%%%%%%%%%%%%%%%%%%%%%%%%%%%%%%%%%%%%%%%%%%%%%%%%

\end{document}

%% file: supp.tex
\onecolumngrid
\makeatletter
\renewcommand{\theequation}{S\arabic{equation}}
\renewcommand{\thefigure}{S\arabic{figure}}
\renewcommand{\thetable}{S\arabic{table}}
\setcounter{secnumdepth}{1}

\begin{center}
{\Large SUPPLEMENTAL MATERIAL}
\end{center}
\begin{center}
\vspace{0.8cm}
{\Large Hybrid sub- and superradiant states in emitter arrays with quantized motion}
\end{center}
\begin{center}
Beatriz Olmos$^1$ and Igor Lesanovsky$^{1,2}$
\end{center}
\begin{center}
$^1${\em Institut f\"ur Theoretische Physik and Center for Integrated Quantum Science and Technology, Universit\"at T\"ubingen, Auf der Morgenstelle 14, 72076 T\"ubingen, Germany}\\
$^2${\em School of Physics and Astronomy and Centre for the Mathematics and Theoretical Physics of Quantum Non-Equilibrium Systems, The University of Nottingham, Nottingham, NG7 2RD, United Kingdom}
\end{center}

\section{Derivation of the Master equation from first principles}

We consider here a system of two-level atoms, whose internal degrees of freedom (separated by an energy $\hbar\omega_0$) interact with the motional ones (vibrations in their dipole traps) and the free radiation field. We start from the Hamiltonian
\begin{eqnarray*}
H=\hbar\sum_{\mathbf{k},\lambda} \nu_kc^\dag_{\mathbf{k}\lambda} c_{\mathbf{k}\lambda}+\hbar\omega_0\sum_j\sigma_j^\dag\sigma_j+\hbar\omega\sum_j a_j^\dag a_j+\hbar\sum_{j,\mathbf{k},\lambda}g_\mathbf{k}^\lambda(\sigma_j^\dag+\sigma_j)(c_{\mathbf{k}\lambda}e^{\mathrm{i}\mathbf{k}\cdot\mathbf{r}_j}+c^\dag_{\mathbf{k}\lambda}e^{-\mathrm{i}\mathbf{k}\cdot\mathbf{r}_j})
\end{eqnarray*}
where we have introduced the the spin ladder operators $\sigma_j=\ket{\downarrow}_j\!\bra{\uparrow}$ and $\sigma_j^\dag=\ket{\uparrow}_j\!\bra{\downarrow}$, the bosonic creation and annihilation operators $c_{\mathbf{k}\lambda}^\dag$ and $c_{\mathbf{k}\lambda}$ of a photon with momentum $\mathbf{k}$, energy $\hbar\nu_k=\hbar c|\mathbf{k}|$ and polarization $\lambda=1,2$, and the coupling constant between the emitters and the radiation field $g^\lambda_\mathbf{k}=\mathbf{d}\cdot\hat{\epsilon}^\lambda_\mathbf{k}\sqrt{\frac{\nu_k}{2\varepsilon_0\hbar V}}$, with $\mathbf{d}$, $\hat{\epsilon}^\lambda_\mathbf{k}$, $V$ being the atomic dipole moment, the field polarization unit vector and the quantization volume, respectively. The position vector $\mathbf{r}_j$ can be written in terms of creation and annihilation operators $a_j^\dagger$ and $a_j$ of a phonon with energy $\hbar\omega\ll\hbar\omega_0$ (representing the quantized motion in the traps) as
\begin{equation*}
    \mathbf{r}_j=\mathbf{r}_j^0+\hat{z}\frac{z_\mathrm{ho}}{\sqrt{2}}(a_j+a^\dag_j),
\end{equation*}
with $z_\mathrm{ho}=\sqrt{\hbar/m\omega}$, $m$ being the mass of the atom. We will work in the Lamb-Dicke regime, where $z_\mathrm{ho}\ll\lambda$, with $\lambda$ being the wavelength of the atomic transition. This allows us to introduce the (small) parameter $\eta=(\mathbf{k}\cdot \hat{z})z_\mathrm{ho}/\sqrt{2}=\ll1$. We go  now into the interaction picture with respect to the atomic and field frequencies, i.e.
\begin{equation*}
    H'=\hbar\omega\sum_j a_j^\dag a_j-\hbar\sum_{j,\mathbf{k},\lambda}g_\mathbf{k}^\lambda(e^{i\omega_0 t}\sigma_j^\dag+e^{-i\omega_0 t}\sigma_j)\left[c_{\mathbf{k}\lambda}e^{\mathrm{i}\eta(a_j^\dag+a_j)}e^{\mathrm{i}(\mathbf{k}\cdot\mathbf{r}_j^0-\nu_kt)}+c^\dag_{\mathbf{k}\lambda}e^{-\mathrm{i}\eta(a_j^\dag+a_j)}e^{-\mathrm{i}(\mathbf{k}\cdot\mathbf{r}_j^0-\nu_kt)}\right].
\end{equation*}
Now we need to go into the interaction picture with respect to the phonons. Here, we perform an expansion for small $\eta$ such that
\begin{equation*}
    e^{\mathrm{i}\omega ta_j^\dag a_j}e^{\mathrm{i}\eta(a_j^\dag+a_j)}e^{-\mathrm{i}\omega ta_j^\dag a_j}\approx1+\mathrm{i}\eta (e^{\mathrm{i}\omega t}a_j^\dag+e^{-\mathrm{i}\omega t}a_j)-\frac{\eta^2}{2} \left[1+2a_j^\dag a_j+e^{2\mathrm{i}\omega t}(a_j^\dag)^2+e^{-2\mathrm{i}\omega t}(a_j)^2\right],
\end{equation*}
which in turn gives the Hamiltonian
\begin{equation*}
    H''=H_0+ H_1+ H_2
\end{equation*}
where
\begin{eqnarray*}
    H_0&=&-\hbar\sum_{j,\mathbf{k},\lambda}g^\lambda_\mathbf{k}A_j(t)\left[c_{\mathbf{k}\lambda}e^{\mathrm{i}(\mathbf{k}\cdot\mathbf{r}_j^0-\nu_kt)}+c_{\mathbf{k}\lambda}^\dag e^{-\mathrm{i}(\mathbf{k}\cdot\mathbf{r}_j^0-\nu_kt)}\right]\\
    H_1&=&-\hbar\sum_{j,\mathbf{k},\lambda}\mathrm{i}\eta g_\mathbf{k}^\lambda A_j(t)\left[c_{\mathbf{k}\lambda}e^{\mathrm{i}(\mathbf{k}\cdot\mathbf{r}_j^0-\nu_kt)}-c_{\mathbf{k}\lambda}^\dag e^{-\mathrm{i}(\mathbf{k}\cdot\mathbf{r}_j^0-\nu_kt)}\right](e^{\mathrm{i}\omega t}a_j^\dag+e^{-\mathrm{i}\omega t}a_j)\\
    H_2&=&\hbar\sum_{j,\mathbf{k},\lambda}\frac{\eta^2}{2}g^\lambda_\mathbf{k}A_j(t)\left[c_{\mathbf{k}\lambda}e^{\mathrm{i}(\mathbf{k}\cdot\mathbf{r}_j^0-\nu_kt)}+c_{\mathbf{k}\lambda}^\dag e^{-\mathrm{i}(\mathbf{k}\cdot\mathbf{r}_j^0-\nu_kt)}\right]\left[1+2a_j^\dag a_j+e^{2\mathrm{i}\omega t}(a_j^\dag)^2+e^{-2\mathrm{i}\omega t}(a_j)^2\right]
\end{eqnarray*}
with
\begin{equation*}
    A_j(t)=e^{i\omega_0 t}\sigma_j^\dag+e^{-i\omega_0 t}\sigma_j.
\end{equation*}
Given the form of the last Hamiltonian, $H_2$, it is convenient to separate two contributions $H_2=H_2^0+H_2^2$, namely the phonon number conserving and non-conserving ones, where
\begin{eqnarray*}
    H_2^0&=&\hbar\sum_{j,\mathbf{k},\lambda}\frac{\eta^2}{2}g_\mathbf{k}^\lambda A_j(t)\left[c_{\mathbf{k}\lambda}e^{\mathrm{i}(\mathbf{k}\cdot\mathbf{r}_j^0-\nu_kt)}+c_{\mathbf{k}\lambda}^\dag e^{-\mathrm{i}(\mathbf{k}\cdot\mathbf{r}_j^0-\nu_kt)}\right]\left[1+2a_j^\dag a_j\right]\\
    H_2^2&=&\hbar\sum_{j,\mathbf{k},\lambda}\frac{\eta^2}{2}g_\mathbf{k}^\lambda A_j(t)\left[c_{\mathbf{k}\lambda}e^{\mathrm{i}(\mathbf{k}\cdot\mathbf{r}_j^0-\nu_kt)}+c_{\mathbf{k}\lambda}^\dag e^{-\mathrm{i}(\mathbf{k}\cdot\mathbf{r}_j^0-\nu_kt)}\right]\left[e^{2\mathrm{i}\omega t}(a_j^\dag)^2+e^{-2\mathrm{i}\omega t}(a_j)^2\right].
\end{eqnarray*}
We are interested here in obtaining the dynamics of the density matrix $\rho$ containing atomic and motional degrees of freedom only, by tracing out the free radiation field. To do so, we introduce our Hamiltonian $H''$ into the Redfield equation
\begin{equation*}
    \dot{\rho}=-\frac{1}{\hbar^2}\int_0^\infty \mathrm{d}\tau \mathrm{Tr}_E\left\{\left[H''(t),\left[H''(t-\tau),\rho(t)\otimes\rho_E\right]\right]\right\},
\end{equation*}
where the subscript $E$ represents the radiation field degrees of freedom. Inspecting now the three terms of the Hamiltonian, we can see that we obtain first a leading contribution, which is exactly the Lindblad master equation for atoms coupled to the radiation field in the absence of atom-phonon coupling \cite{Dicke1954,Lehmberg1970,James1993}. Next, the terms proportional to $\eta$ contain $\left[H_0(t),\left[H_1(t-\tau),\rho\right]\right]$ and $\left[H_1(t),\left[H_0(t-\tau),\rho\right]\right]$. The contribution of these terms is negligible under the secular (rotating wave) approximation, as here all terms are proportional to $e^{\pm\mathrm{i}\omega t}$. The next and final order we wish to consider is the proportional to $\eta^2$. Here, there are two types of contributions we have to consider: the ones that arise from $\left[H_1(t),\left[H_1(t-\tau),\rho\right]\right]$ and the ones with $\left[H_0(t),\left[H_2^0(t-\tau),\rho\right]\right]$ and $\left[H_2^0(t),\left[H_0(t-\tau),\rho\right]\right]$. We will calculate these terms explicitly. Note, that we neglect directly the terms involving the phonon number non-conserving Hamiltonian $H_2^2$, as again they drop out under the secular approximation. In the following, we explicitly calculate all on these surviving contributions.

\subsection{Leading term: $\left[H_0(t),\left[H_0(t-\tau),\rho\right]\right]$}

\begin{figure}
    \centering
    \includegraphics[width=0.8\columnwidth]{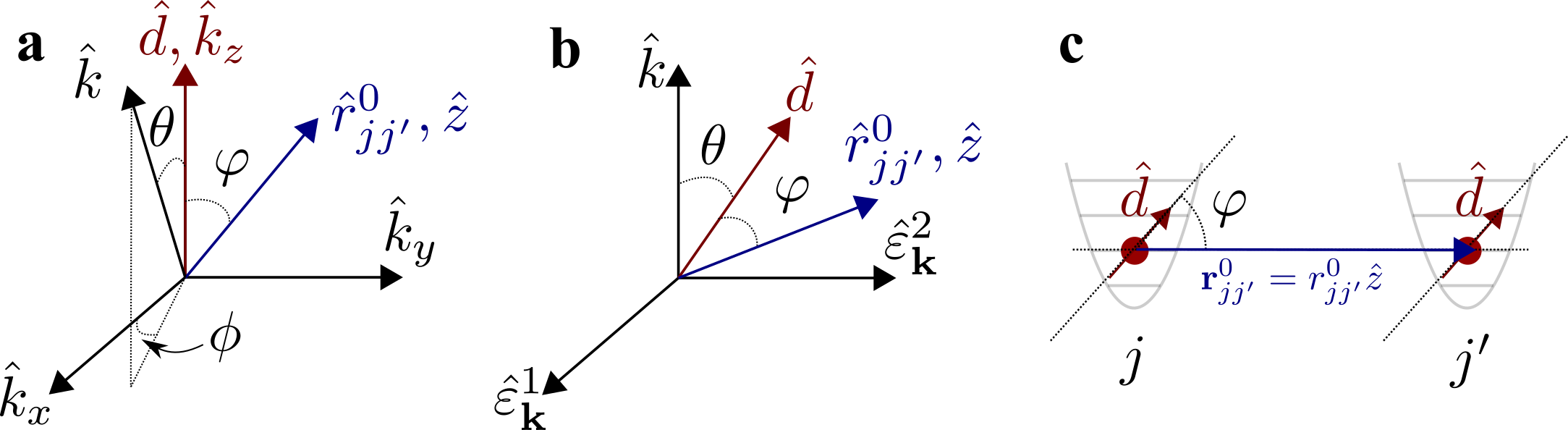}
    \caption{\textbf{a:} Spherical coordinates of the unit vector $\hat{k}$ in $\mathbf{k}$-space. \textbf{b:} The two field polarizations $\varepsilon_\mathbf{k}^\lambda$ with $\lambda=1,2$ and the unit vector $\hat{k}$ form a coordinate system, such that the unit dipole moment $\hat{d}$ satisfies $(\hat{d}\cdot\hat{k})^2+(\hat{d}\cdot\hat{\varepsilon}_\mathbf{k}^1)^2+(\hat{d}\cdot\hat{\varepsilon}_\mathbf{k}^2)^2=1$. \textbf{c:} The dipole moment $\mathbf{d}$ of the atomic transition forms an angle $\varphi$ with respect to $\mathbf{r}_{jj'}^0$, i.e., the vector that separates the atoms $j$ and $j'$, which is parallel to the $z$-axis.}
    \label{fig:figSM}
\end{figure}

Starting from the Redfield equation, and introducing the notation
\begin{equation*}
    B_{\lambda\mathbf{k}j}(t)=c_{\mathbf{k}\lambda}e^{\mathrm{i}(\mathbf{k}\cdot\mathbf{r}_j^0-\nu_kt)},
\end{equation*}
we have
\begin{equation*}
    \begin{split}
    \dot{\rho}=&-\frac{1}{\hbar^2}\int_0^\infty \mathrm{d}\tau \mathrm{Tr}_E\left\{\left[H_0(t),\left[H_0(t-\tau),\rho(t)\otimes\rho_E\right]\right]\right\}\\
    =&-\!\!\!\!\!\!\sum_{\mathbf{k}\mathbf{k}'jj'\lambda\lambda'}g_\mathbf{k}^\lambda {g_{\mathbf{k}'}^{\lambda'}}^*\int_0^\infty \mathrm{d}\tau\mathrm{Tr}_E\left\{\left[A_j(t)\left(B_{\lambda\mathbf{k}j}(t)+B^\dag_{\lambda\mathbf{k}j}(t)\right),\left[A_{j'}(t-\tau)\left(B_{\lambda'\mathbf{k}'j'}(t-\tau)+B^\dag_{\lambda'\mathbf{k}'j'}(t-\tau)\right),\rho(t)\otimes\rho_E\right]\right]\right\}\\
    =&-\!\!\!\!\!\!\sum_{\mathbf{k}\mathbf{k}'jj'\lambda\lambda'}g^\lambda_\mathbf{k}{g^{\lambda'}_{\mathbf{k}'}}^*\int_0^\infty \mathrm{d}\tau\mathrm{Tr}_E\left\{\left[A_j(t)\left(B_{\lambda\mathbf{k}j}(t)+B^\dag_{\lambda\mathbf{k}j}(t)\right),A_{j'}(t-\tau)\left(B_{\lambda'\mathbf{k}'j'}(t-\tau)+B^\dag_{\lambda'\mathbf{k}'j'}(t-\tau)\right)\rho(t)\otimes\rho_E\right]\right.\\
    &\left.-\left[A_j(t)\left(B_{\lambda\mathbf{k}j}(t)+B^\dag_{\lambda\mathbf{k}j}(t)\right),\rho(t)\otimes\rho_EA_{j'}(t-\tau)\left(B_{\lambda'\mathbf{k}'j'}(t-\tau)+B^\dag_{\lambda'\mathbf{k}'j'}(t-\tau)\right)\right]\right\}\\
    =&-\!\!\!\!\!\!\sum_{\mathbf{k}\mathbf{k}'jj'\lambda\lambda'}g^\lambda_\mathbf{k}{g^{\lambda'}_{\mathbf{k}'}}^*\int_0^\infty \mathrm{d}\tau\mathrm{Tr}_E\left\{A_j(t)A_{j'}(t-\tau)\left(B_{\lambda\mathbf{k}j}(t)+B^\dag_{\lambda\mathbf{k}j}(t)\right)\left(B_{\lambda'\mathbf{k}'j'}(t-\tau)+B^\dag_{\lambda'\mathbf{k}'j'}(t-\tau)\right)\rho(t)\otimes\rho_E\right.\\
    &-A_{j'}(t-\tau)\left(B_{\lambda'\mathbf{k}'j'}(t-\tau)+B^\dag_{\lambda'\mathbf{k}'j'}(t-\tau)\right)\rho(t)\otimes\rho_EA_j(t)\left(B_{\lambda\mathbf{k}j}(t)+B^\dag_{\lambda\mathbf{k}j}(t)\right)\\
    &-A_j(t)\left(B_{\lambda\mathbf{k}j}(t)+B^\dag_{\lambda\mathbf{k}j}(t)\right)\rho(t)\otimes\rho_EA_{j'}(t-\tau)\left(B_{\lambda'\mathbf{k}'j'}(t-\tau)+B^\dag_{\lambda'\mathbf{k}'j'}(t-\tau)\right)\\
    &\left.+\rho(t)\otimes\rho_EA_{j'}(t-\tau)\left(B_{\lambda'\mathbf{k}'j'}(t-\tau)+B^\dag_{\lambda'\mathbf{k}'j'}(t-\tau)\right)A_j(t)\left(B_{\lambda\mathbf{k}j}(t)+B^\dag_{\lambda\mathbf{k}j}(t)\right)\right\}.
    \end{split}
\end{equation*}
Here, we can eliminate the majority of terms by considering the environment as a zero-temperature photon bath. Thus,
\begin{eqnarray*}
\left<B_{\lambda\mathbf{k}j}(t)B^\dag_{\lambda'\mathbf{k}'j'}(t')\right>=\mathrm{Tr}_E\left\{B_{\lambda\mathbf{k}j}(t)B^\dag_{\lambda'\mathbf{k}'j'}(t')\rho_E\right\}=\delta_{\mathbf{k}\mathbf{k}'}\delta_{\lambda\lambda'}e^{\mathrm{i}\left[\mathbf{k}\cdot\mathbf{r}^0_{jj'}-\nu_k(t-t')\right]}
\end{eqnarray*}
with $\mathbf{r}^0_{jj'}=\mathbf{r}^0_{j}-\mathbf{r}^0_{j'}$, are the only non-zero expectation values with respect to the environment. The equation is then reduced to
\begin{equation*}
    \begin{split}
    \dot{\rho}=&-\sum_{\mathbf{k}\lambda jj'}|g^\lambda_\mathbf{k}|^2\int_0^\infty \mathrm{d}\tau\left\{\left[A_j(t)A_{j'}(t-\tau)\rho(t)-A_{j'}(t-\tau)\rho(t)A_j(t)\right]e^{\mathrm{i}\left(\mathbf{k}\cdot\mathbf{r}^0_{jj'}-\nu_k\tau\right)}\right.\\
    &+\left.\left[\rho(t)A_{j'}(t-\tau)A_j(t)-A_j(t)\rho(t)A_{j'}(t-\tau)\right]e^{-\mathrm{i}\left(\mathbf{k}\cdot \mathbf{r}^0_{jj'}-\nu_k\tau\right)}\right\}.
    \end{split}
\end{equation*}
Now we have to write out the $A_j(t)$ terms and perform the secular approximation, neglecting all terms that still contain an imaginary exponential that depends on $t$ (also called the rotating wave approximation). For example:
\begin{equation*}
    A_j(t)A_{j'}(t-\tau)\rho(t)=(e^{i\omega_0 t}\sigma_j^\dag+e^{-i\omega_0 t}\sigma_j) (e^{i\omega_0 t}e^{-i\omega_0 \tau}\sigma_{j'}^\dag+e^{-i\omega_0 t}e^{i\omega_0 \tau}\sigma_{j'})\rho\approx (\sigma_j^\dag \sigma_{j'} e^{i\omega_0 \tau}+\sigma_j \sigma_{j'}^\dag e^{-i\omega_0 \tau})\rho.
\end{equation*}
After this, we have
\begin{equation*}
    \begin{split}
    \dot{\rho}=&-\sum_{\mathbf{k}\lambda jj'}|g^\lambda_\mathbf{k}|^2\int_0^\infty \mathrm{d}\tau\left\{\left[(\sigma_j^\dag \sigma_{j'}\rho-\sigma_{j'}\rho\sigma_j^\dag) e^{i(\omega_0-\nu_k) \tau}+(\sigma_j \sigma_{j'}^\dag\rho -\sigma_{j'}^\dag\rho\sigma_j)e^{-i(\omega_0+\nu_k)\tau}\right]e^{\mathrm{i}\mathbf{k}\cdot\mathbf{r}^0_{jj'}}+\right.\\
    &\left.\left[(\rho\sigma_{j'}\sigma_j^\dag-\sigma_j^\dag\rho\sigma_{j'})e^{i(\omega_0+\nu_k)\tau}+(\rho\sigma_{j'}^\dag\sigma_j -\sigma_j\rho\sigma_{j'}^\dag) e^{-i(\omega_0-\nu_k)\tau}\right]e^{-\mathrm{i}\mathbf{k}\cdot\mathbf{r}^0_{jj'}}\right\}\\
    =&-\sum_{\mathbf{k}\lambda jj'}|g^\lambda_\mathbf{k}|^2e^{\mathrm{i}\mathbf{k}\cdot\mathbf{r}^0_{jj'}}\int_0^\infty \mathrm{d}\tau\left[(\sigma_j^\dag \sigma_{j'}\rho-\sigma_{j'}\rho\sigma_j^\dag) e^{i(\omega_0-\nu_k) \tau}+(\sigma_j \sigma_{j'}^\dag\rho -\sigma_{j'}^\dag\rho\sigma_j)e^{-i(\omega_0+\nu_k)\tau}\right.\\
    &\left.+(\rho\sigma_{j}\sigma_{j'}^\dag-\sigma_{j'}^\dag\rho\sigma_{j})e^{i(\omega_0+\nu_k)\tau}+(\rho\sigma_{j}^\dag\sigma_{j'} -\sigma_{j'}\rho\sigma_{j}^\dag) e^{-i(\omega_0-\nu_k)\tau}\right],
    \end{split}
\end{equation*}
where in the last step we have exchanged $j\to j'$ in the second line. We perform the time integral, which is done by using the so-called Heitler function
\begin{eqnarray*}
\int_0^\infty d\tau\, e^{-i(\omega_0 \pm \nu_k)\tau} &=& \pi \delta(\omega_0 \pm \nu_k) - i \mathcal{P} \Big(\frac{1}{\omega_0 \pm \nu_k}\Big)\\
\int_0^\infty d\tau\, e^{i(\omega_0 \pm \nu_k)\tau} &=& \pi \delta(\omega_0 \pm \nu_k) + i \mathcal{P} \Big(\frac{1}{\omega_0 \pm \nu_k}\Big),
\end{eqnarray*}
where ${\cal P}$ represents the Principal Cauchy Value. After this, we can separate two types of contributions, the ones proportional to the delta function and the Principal Cauchy Value. Moreover, we neglect all the terms proportional to $\delta(\omega_0+\nu_k)$, which will lead to a zero contribution after performing the integral in $k$-space, which runs for positive frequencies only. After this, we obtain
\begin{equation*}
    \begin{split}
    \dot{\rho}=&-\sum_{\mathbf{k}\lambda jj'}|g^\lambda_\mathbf{k}|^2e^{\mathrm{i}\mathbf{k}\cdot\mathbf{r}^0_{jj'}}\left[\pi \delta(\omega_0 - \nu_k)\left(\sigma_j^\dag \sigma_{j'}\rho+\rho\sigma_{j}^\dag\sigma_{j'} -2\sigma_{j'}\rho\sigma_{j}^\dag\right)\right.\\
    &+\mathrm{i}\left[(\sigma_j^\dag \sigma_{j'}\rho-\rho\sigma_{j}^\dag\sigma_{j'}) \mathcal{P} \Big(\frac{1}{\omega_0 - \nu_k}\Big) -(\sigma_j \sigma_{j'}^\dag\rho -\rho\sigma_{j}\sigma_{j'}^\dag)\mathcal{P} \Big(\frac{1}{\omega_0 + \nu_k}\Big)\right],
    \end{split}
\end{equation*}
such that we have a real and an imaginary contribution. In order continue we need first to turn the sum over $\mathbf{k}$ into an integral, yielding,
\begin{equation*}
    \sum_\mathbf{k}\to\frac{V}{(2\pi)^3}\sum_{\lambda=1,2}\int_{-\infty}^\infty\mathrm{d}k_x\int_{-\infty}^\infty\mathrm{d}k_y\int_{-\infty}^\infty\mathrm{d}k_z=\frac{V}{(2\pi c)^3}\sum_{\lambda=1,2}\int_0^{2\pi}\mathrm{d}\phi\int_0^\pi\mathrm{d}\theta \sin\theta\int_0^\infty \mathrm{d}\nu_k\nu_k^2,
\end{equation*}
where we have used spherical coordinates (see Fig. \ref{fig:figSM}a). Writing the dipole moment in terms of the perpendicular vectors $\varepsilon_\mathbf{k}^\lambda$ with $\lambda=1,2$ and $\hat{k}$, the sum over the polarizations yields
\begin{equation*}
    \sum_{\lambda=1,2}|\hat{\varepsilon}^\lambda_\mathbf{k}\cdot\hat{d}|^2=1-|\hat{k}\cdot\hat{d}|^2=\sin^2\theta,
\end{equation*}
see Figure \ref{fig:figSM}b.
Introducing this into the master equation, we have
\begin{equation*}
    \begin{split}
    \dot{\rho}=&\frac{\omega_0^3 |\mathbf{d}|^2\pi}{\epsilon_0\hbar(2\pi c )^3}\sum_{jj'}\left(\sigma_{j'}\rho\sigma_{j}^\dag-\frac{1}{2}\left\{\sigma_j^\dag \sigma_{j'},\rho\right\} \right)\int_0^{2\pi}\mathrm{d}\phi\int_0^\pi\mathrm{d}\theta \sin^3\theta e^{\mathrm{i}\frac{\omega_0}{c}\hat{k}\cdot\mathbf{r}^0_{jj'}}\\
    &-\frac{\mathrm{i}|\mathbf{d}|^2}{2\epsilon_0\hbar(2\pi c )^3}\sum_{jj'}\int_0^\infty \mathrm{d}\nu_k\nu_k^3\int_0^{2\pi}\mathrm{d}\phi\int_0^\pi\mathrm{d}\theta \sin^3\theta e^{\mathrm{i}\mathbf{k}\cdot\mathbf{r}^0_{jj'}}\left[(\sigma_j^\dag \sigma_{j'}\rho-\rho\sigma_{j}^\dag\sigma_{j'}) \mathcal{P} \Big(\frac{1}{\omega_0 - \nu_k}\Big)\right.\\
    &\left.-(\sigma_j \sigma_{j'}^\dag\rho -\rho\sigma_{j}\sigma_{j'}^\dag)\mathcal{P} \Big(\frac{1}{\omega_0 + \nu_k}\Big)\right],
    \end{split}
\end{equation*}
and, considering that the integral over the angular part in $\mathbf{k}$-space is given by
\begin{equation*}
    \int_0^{2\pi}\mathrm{d}\phi\int_0^\pi\mathrm{d}\theta \sin^3\theta e^{\mathrm{i}\mathbf{k}\cdot\mathbf{r}^0_{jj'}}=4\pi F(kr^0_{jj'}),
\end{equation*}
with
\begin{equation*}
    F(kr^0_{jj'})=\left[1-(\hat{d}\cdot\hat{r}^0_{jj'})^2\right]\frac{\sin{kr^0_{jj'}}}{kr^0_{jj'}}+\left[1-3(\hat{d}\cdot\hat{r}^0_{jj'})^2\right]\left[\frac{\cos{kr^0_{jj'}}}{(kr^0_{jj'})^2}-\frac{\sin{kr^0_{jj'}}}{(kr^0_{jj'})^3}\right],
\end{equation*}
with $\hat{d}\cdot\hat{r}^0_{jj'}=\cos\varphi$ (see Fig. \ref{fig:figSM}c), we obtain
\begin{equation*}
    \begin{split}
    \dot{\rho}=&\frac{\omega_0^3 |\mathbf{d}|^2}{3\epsilon_0\hbar\pi c^3}\sum_{jj'}\frac{3}{2}F(k_0r^0_{jj'})\left(\sigma_{j'}\rho\sigma_{j}^\dag-\frac{1}{2}\left\{\sigma_j^\dag \sigma_{j'},\rho\right\} \right)-\frac{\mathrm{i}|\mathbf{d}|^2}{3\epsilon_0\hbar c^3\pi^2}\sum_{jj'}\int_0^\infty \mathrm{d}\nu_k\nu_k^3 \frac{3}{4}F\left(\frac{\nu_k}{c}r^0_{jj'}\right)\\
    &\times\left[(\sigma_j^\dag \sigma_{j'}\rho-\rho\sigma_{j}^\dag\sigma_{j'}) \mathcal{P} \Big(\frac{1}{\omega_0 - \nu_k}\Big) -(\sigma_j \sigma_{j'}^\dag\rho -\rho\sigma_{j}\sigma_{j'}^\dag)\mathcal{P} \Big(\frac{1}{\omega_0 + \nu_k}\Big)\right].
    \end{split}
\end{equation*}
Now we use that $F$ is invariant under the exchange $j\to j'$ and obtain
\begin{equation*}
    \begin{split}
    \dot{\rho}=&\frac{\omega_0^3 |\mathbf{d}|^2}{3\epsilon_0\hbar\pi c^3}\sum_{jj'}\frac{3}{2}F(k_0r^0_{jj'})\left(\sigma_{j'}\rho\sigma_{j}^\dag-\frac{1}{2}\left\{\sigma_j^\dag \sigma_{j'},\rho\right\} \right)\\
    &+\frac{\mathrm{i}|\mathbf{d}|^2}{3\epsilon_0\hbar c^3\pi^2}\sum_{j\neq j'}\left[\sigma_{j'}^\dag\sigma_j,\rho\right]\frac{3}{4}\mathcal{P}\int_0^\infty \mathrm{d}\nu_k\nu_k^3 F\left(\frac{\nu_k}{c}r^0_{jj'}\right)\left[ \frac{1}{\nu_k+\omega_0} +\frac{1}{ \nu_k-\omega_0}\right]\\
    &+\frac{\mathrm{i}|\mathbf{d}|^2}{3\epsilon_0\hbar c^3\pi^2}\frac{3}{4}\sum_{j}\mathcal{P}\int_0^\infty \mathrm{d}\nu_k\nu_k^3 \left[ \frac{\rho\sigma_z^j}{\nu_k-\omega_0} -\frac{\sigma_z^j\rho}{\nu_k+\omega_0}\right].
    \end{split}
\end{equation*}
The last term has the same value for all atoms and is absorbed into the atomic energy \cite{Lehmberg1970}. To evaluate the integral in the second term, we use that the function $F$ is even in the variable $\nu_k$, such that we can extend the integral and obtain
\begin{equation*}
    \mathcal{P}\int_0^\infty \mathrm{d}\nu_k\nu_k^3 F\left(\frac{\nu_k}{c}r^0_{jj'}\right)\left[ \frac{1}{\nu_k+\omega_0} +\frac{1}{ \nu_k-\omega_0}\right]=\mathcal{P}\int^\infty_{-\infty} \mathrm{d}\nu_k F\left(\frac{\nu_k}{c}r^0_{jj'}\right) \frac{\nu_k^3}{\nu_k-\omega_0}=\pi \omega_0^3  G\left(k_0r^0_{jj'}\right),
\end{equation*}
with
\begin{equation*}
    G(k_0r^0_{jj'})=\left[1-(\hat{d}\cdot\hat{r}^0_{jj'})^2\right]\frac{\cos{k_0r^0_{jj'}}}{k_0r^0_{jj'}}-\left[1-3(\hat{d}\cdot\hat{r}^0_{jj'})^2\right]\left[\frac{\sin{k_0r^0_{jj'}}}{(k_0r^0_{jj'})^2}+\frac{\cos{k_0r^0_{jj'}}}{(k_0r^0_{jj'})^3}\right].
\end{equation*}
Finally, making the definitions
\begin{equation*}
    \gamma=\frac{\omega_0^3 |\mathbf{d}|^2}{3\epsilon_0\hbar\pi c^3}\qquad \Gamma_{jj'}=\frac{3}{2}\gamma F(k_0r^0_{jj'})\qquad V_{jj'}=\frac{3}{4}\gamma G(k_0r^0_{jj'}),
\end{equation*}
we obtain
\begin{equation*}
    \dot{\rho}=\sum_{jj'}\Gamma_{jj'}\left(\sigma_{j'}\rho\sigma_{j}^\dag-\frac{1}{2}\left\{\sigma_j^\dag \sigma_{j'},\rho\right\} \right)+\mathrm{i}\sum_{j\neq j'}V_{jj'}\left[\sigma_{j'}^\dag\sigma_j,\rho\right],
\end{equation*}
which is the zero-th order for the master equation. We will now apply these same steps to the other terms that constitute the second order perturbation.

\subsection{Terms proportional to $\eta^2$: $\left[H_1(t),\left[H_1(t-\tau),\rho\right]\right]$}

Starting again from the Redfield equation, and introducing the notation
\begin{equation*}
    C_j(t)=e^{\mathrm{i}\omega t}a_j^\dag+e^{-\mathrm{i}\omega t}a_j, \qquad \tilde{A}_j(t)=A_j(t)C_j(t)
\end{equation*}
we have
\begin{equation*}
    \begin{split}
    \dot{\rho}=&-\frac{1}{\hbar^2}\int_0^\infty \mathrm{d}\tau \mathrm{Tr}_E\left\{\left[H_1(t),\left[H_1(t-\tau),\rho(t)\otimes\rho_E\right]\right]\right\}
    \\=&\!\!\!\!\!\!\sum_{\mathbf{k}\mathbf{k}'jj'\lambda\lambda'}\eta^2g^\lambda_\mathbf{k}{g^{\lambda'}_{\mathbf{k}'}}^*\int_0^\infty\mathrm{d}\tau\mathrm{Tr}_E\left\{\left[\tilde{A}_j(t)\left(B_{\lambda\mathbf{k}j}(t)-B^\dag_{\lambda\mathbf{k}j}(t)\right),\left[\tilde{A}_{j'}(t-\tau)\left(B_{\lambda'\mathbf{k}'j'}(t-\tau)-B^\dag_{\lambda'\mathbf{k}'j'}(t-\tau)\right),\rho(t)\otimes\rho_E\right]\right]\right\}\\
    =&\!\!\!\!\!\! \sum_{\mathbf{k}\mathbf{k}'jj'\lambda\lambda'}\eta^2g^\lambda_\mathbf{k}{g^{\lambda'}_{\mathbf{k}'}}^*\int_0^\infty \mathrm{d}\tau\mathrm{Tr}_E\left\{\left[\tilde{A}_j(t)\left(B_{\lambda\mathbf{k}j}(t)-B^\dag_{\lambda\mathbf{k}j}(t)\right),\tilde{A}_{j'}(t-\tau)\left(B_{\lambda'\mathbf{k}'j'}(t-\tau)-B^\dag_{\lambda'\mathbf{k}'j'}(t-\tau)\right)\rho(t)\otimes\rho_E\right]\right.\\
    &\left.-\left[\tilde{A}_j(t)\left(B_{\lambda\mathbf{k}j}(t)-B^\dag_{\lambda\mathbf{k}j}(t)\right),\rho(t)\otimes\rho_E\tilde{A}_{j'}(t-\tau)\left(\lambda'B_{\mathbf{k}'j'}(t-\tau)-B^\dag_{\lambda'\mathbf{k}'j'}(t-\tau)\right)\right]\right\}\\=\!\!\!\!\!\!&\sum_{\mathbf{k}\mathbf{k}'jj'\lambda\lambda'}\eta^2g^\lambda_\mathbf{k}{g^{\lambda'}_{\mathbf{k}'}}^*\int_0^\infty \mathrm{d}\tau\mathrm{Tr}_E\left\{\tilde{A}_j(t)\tilde{A}_{j'}(t-\tau)\left(B_{\lambda\mathbf{k}j}(t)-B^\dag_{\lambda\mathbf{k}j}(t)\right)\left(B_{\lambda'\mathbf{k}'j'}(t-\tau)-B^\dag_{\lambda'\mathbf{k}'j'}(t-\tau)\right)\rho(t)\otimes\rho_E\right.\\
    &-\tilde{A}_{j'}(t-\tau)\left(B_{\lambda'\mathbf{k}'j'}(t-\tau)-B^\dag_{\lambda'\mathbf{k}'j'}(t-\tau)\right)\rho(t)\otimes\rho_E\tilde{A}_j(t)\left(B_{\lambda\mathbf{k}j}(t)-B^\dag_{\lambda\mathbf{k}j}(t)\right)\\
    &-\tilde{A}_j(t)\left(B_{\lambda\mathbf{k}j}(t)-B^\dag_{\lambda\mathbf{k}j}(t)\right)\rho(t)\otimes\rho_E\tilde{A}_{j'}(t-\tau)\left(B_{\lambda'\mathbf{k}'j'}(t-\tau)-B^\dag_{\lambda'\mathbf{k}'j'}(t-\tau)\right)\\
    &\left.+\rho(t)\otimes\rho_E\tilde{A}_{j'}(t-\tau)\left(B_{\lambda'\mathbf{k}'j'}(t-\tau)-B^\dag_{\lambda'\mathbf{k}'j'}(t-\tau)\right)\tilde{A}_j(t)\left(B_{\lambda\mathbf{k}j}(t)-B^\dag_{\lambda\mathbf{k}j}(t)\right)\right\}.
\end{split}
\end{equation*}
Taking again the expectation value with respect to the environment, i.e. the radiation field at zero temperature, we obtain
\begin{equation*}
    \begin{split}
    \dot{\rho}=&-\sum_{\mathbf{k}\lambda jj'}\eta^2|g^\lambda_\mathbf{k}|^2\int_0^\infty \mathrm{d}\tau\left\{\left[\tilde{A}_j(t)\tilde{A}_{j'}(t-\tau)\rho(t)-\tilde{A}_{j'}(t-\tau)\rho(t)\tilde{A}_j(t)\right]e^{\mathrm{i}\left(\mathbf{k}\cdot\mathbf{r}^0_{jj'}-\nu_k\tau\right)}\right.\\
    &+\left.\left[\rho(t)\tilde{A}_{j'}(t-\tau)\tilde{A}_j(t)-\tilde{A}_j(t)\rho(t)\tilde{A}_{j'}(t-\tau)\right]e^{-\mathrm{i}\left(\mathbf{k}\cdot\mathbf{r}^0_{jj'}-\nu_k\tau\right)}\right\}.
\end{split}
\end{equation*}
We write out the $\tilde{A}_j(t)$ terms and perform the secular approximation, neglecting all terms that contain a exponential function that depends on the time $t$, for example,
\begin{equation*}
    \begin{split}
    \tilde{A}_j(t)\tilde{A}_{j'}(t-\tau)\rho(t)=&(e^{i\omega_0 t}\sigma_j^\dag+e^{-i\omega_0 t}\sigma_j)(e^{\mathrm{i}\omega t}a_j^\dag+e^{-\mathrm{i}\omega t}a_j)(e^{i\omega_0 t}e^{-i\omega_0 \tau}\sigma_{j'}^\dag+e^{-i\omega_0 t}e^{i\omega_0 \tau}\sigma_{j'})(e^{\mathrm{i}\omega t}e^{-\mathrm{i}\omega \tau}a_{j'}^\dag+e^{-\mathrm{i}\omega t}e^{\mathrm{i}\omega \tau}a_{j'})\rho\\
    \approx& (\sigma_j^\dag a_j^\dag \sigma_{j'}a_{j'} e^{i(\omega_0+\omega )\tau}+\sigma_j^\dag a_j \sigma_{j'}a_{j'}^\dag e^{i(\omega_0-\omega )\tau}+\sigma_j a_j \sigma_{j'}^\dag a_{j'}^\dag e^{-i(\omega_0+\omega) \tau}+\sigma_j a_j^\dag \sigma_{j'}^\dag a_{j'} e^{-i(\omega_0-\omega) \tau})\rho.
\end{split}
\end{equation*}
Now we can write, using the shortcut notation $\omega_\pm=\omega_0\pm\omega$ and $J_j=\sigma_j a_j$ and $L_j=\sigma_j a_j^\dag$,
\begin{equation*}
    \begin{split}
    \dot{\rho}=&-\sum_{\mathbf{k}\lambda jj'}\eta^2|g^\lambda_\mathbf{k}|^2\int_0^\infty \mathrm{d}\tau\left\{\left[(J_j^\dag  J_{j'}\rho-J_{j'}\rho J_j^\dag) e^{i(\omega_+-\nu_k) \tau}+(J_j J_{j'}^\dag\rho -J_{j'}^\dag\rho J_j)e^{-i(\omega_++\nu_k)\tau}\right]e^{\mathrm{i}\mathbf{k}\cdot\mathbf{r}^0_{jj'}}\right.\\
    &+\left.\left[(\rho J_{j'}^\dag J_j-J_j^\dag\rho J_{j'})e^{i(\omega_++\nu_k)\tau}+(\rho J_{j'}J_j^\dag -J_j\rho J_{j'}^\dag) e^{-i(\omega_+-\nu_k)\tau}\right]e^{-\mathrm{i}\mathbf{k}\cdot\mathbf{r}^0_{jj'}}\right\}\\
    &-\sum_{\mathbf{k}\lambda jj'}\eta^2|g^\lambda_\mathbf{k}|^2\int_0^\infty \mathrm{d}\tau\left\{\left[(L_j^\dag  L_{j'}\rho-L_{j'}\rho L_j^\dag) e^{i(\omega_--\nu_k) \tau}+(L_j L_{j'}^\dag\rho -L_{j'}^\dag\rho L_j)e^{-i(\omega_-+\nu_k)\tau}\right]e^{\mathrm{i}\mathbf{k}\cdot\mathbf{r}^0_{jj'}}\right.\\
    &+\left.\left[(\rho L_{j'}^\dag L_j-L_j^\dag\rho L_{j'})e^{i(\omega_-+\nu_k)\tau}+(\rho L_{j'}L_j^\dag -L_j\rho L_{j'}^\dag) e^{-i(\omega_--\nu_k)\tau}\right]e^{-\mathrm{i}\mathbf{k}\cdot\mathbf{r}^0_{jj'}}\right\}\\
    =&-\sum_{\mathbf{k}\lambda jj'}\eta^2|g^\lambda_\mathbf{k}|^2e^{\mathrm{i}\mathbf{k}\cdot\mathbf{r}^0_{jj'}}\int_0^\infty \mathrm{d}\tau\left[(J_j^\dag  J_{j'}\rho-J_{j'}\rho J_j^\dag) e^{i(\omega_+-\nu_k) \tau}+(J_j J_{j'}^\dag\rho -J_{j'}^\dag\rho J_j)e^{-i(\omega_++\nu_k)\tau}\right.+\\
    &\left.(\rho J_{j}^\dag J_{j'}-J_{j'}^\dag\rho J_{j})e^{i(\omega_++\nu_k)\tau}+(\rho J_{j}J_{j'}^\dag -J_{j'}\rho J_{j}^\dag) e^{-i(\omega_+-\nu_k)\tau}\right]\\
    &-\sum_{\mathbf{k}\lambda jj'}\eta^2|g^\lambda_\mathbf{k}|^2e^{\mathrm{i}\mathbf{k}\cdot\mathbf{r}^0_{jj'}}\int_0^\infty \mathrm{d}\tau\left[(L_j^\dag  L_{j'}\rho-L_{j'}\rho L_j^\dag) e^{i(\omega_--\nu_k) \tau}+(L_j L_{j'}^\dag\rho -L_{j'}^\dag\rho L_j)e^{-i(\omega_-+\nu_k)\tau}+\right.\\
    &\left.(\rho L_{j}^\dag L_{j'}-L_{j'}^\dag\rho L_{j})e^{i(\omega_-+\nu_k)\tau}+(\rho L_{j}L_{j'}^\dag -L_{j'}\rho L_{j}^\dag) e^{-i(\omega_--\nu_k)\tau}\right],
\end{split}
\end{equation*}
where in the last step we have exchanged $j\to j'$ in the second lines. We can see now that we have the same structure as in the leading case, but with two terms with two frequencies and two jump operators:
\begin{equation*}
    J_j=\sigma_ja_j \quad \text{with}\quad \omega_+\quad\text{and}\quad L_j=\sigma_ja^\dag_j \quad \text{with}\quad \omega_-.
\end{equation*}
All of the mathematical steps that follow this point are thus parallel to the ones we used for the leading order case. We use first the Heitler function and obtain
\begin{equation*}
    \begin{split}
    \dot{\rho}=&-\sum_{\mathbf{k}\lambda jj'}\eta^2|g^\lambda_\mathbf{k}|^2e^{\mathrm{i}\mathbf{k}\cdot\mathbf{r}^0_{jj'}}\left[\pi \delta(\omega_+ - \nu_k)\left(J_j^\dag J_{j'}\rho+\rho J_{j}^\dag J_{j'} -2J_{j'}\rho J_{j}^\dag\right)\right.\\
    &+\mathrm{i}\left[(J_j^\dag J_{j'}\rho-\rho J_{j}^\dag J_{j'}) \mathcal{P} \Big(\frac{1}{\omega_+ - \nu_k}\Big) -(J_j J_{j'}^\dag\rho -\rho J_{j}J_{j'}^\dag)\mathcal{P} \Big(\frac{1}{\omega_+ + \nu_k}\Big)\right]\\
    &-\sum_{\mathbf{k}\lambda jj'}\eta^2|g^\lambda_\mathbf{k}|^2e^{\mathrm{i}\mathbf{k}\cdot\mathbf{r}^0_{jj'}}\left[\pi \delta(\omega_- - \nu_k)\left(L_j^\dag L_{j'}\rho+\rho L_{j}^\dag L_{j'} -2L_{j'}\rho L_{j}^\dag\right)\right.\\
    &+\mathrm{i}\left[(L_j^\dag L_{j'}\rho-\rho L_{j}^\dag L_{j'}) \mathcal{P} \Big(\frac{1}{\omega_- - \nu_k}\Big) -(L_j L_{j'}^\dag\rho -\rho L_{j}L_{j'}^\dag)\mathcal{P} \Big(\frac{1}{\omega_- + \nu_k}\Big)\right].
\end{split}
\end{equation*}
Taking again the continuum limit of the momentum sum, and realizing the sum over the two polarizations, we obtain, after substituting $\eta^2=(\mathbf{k}\cdot\hat{z})^2 z_\mathrm{ho}^2/2$, that
\begin{equation*}
    \begin{split}
    \dot{\rho}=&\frac{\omega_+^3 z_\mathrm{ho}^2 |\mathbf{d}|^2\pi}{2\epsilon_0\hbar(2\pi)^3 c ^3}\sum_{jj'}\left(J_{j'}\rho J_{j}^\dag-\frac{1}{2}\left\{J_j^\dag J_{j'},\rho\right\} \right)\int_0^{2\pi}\mathrm{d}\phi\int_0^\pi\mathrm{d}\theta \sin^3\theta(\mathbf{k_+}\cdot\hat{z})^2 e^{\mathrm{i}\frac{\omega_+}{c}\hat{k}\cdot\mathbf{r}^0_{jj'}}\\
    &-\frac{\mathrm{i}|\mathbf{d}|^2z_\mathrm{ho}^2}{4\epsilon_0\hbar(2\pi)^3 c^3}\sum_{jj'}\int_0^\infty \mathrm{d}\nu_k\nu_k^3\int_0^{2\pi}\mathrm{d}\phi\int_0^\pi\mathrm{d}\theta \sin^3\theta (\mathbf{k}\cdot\hat{z})^2e^{\mathrm{i}\mathbf{k}\cdot\mathbf{r}^0_{jj'}}\\
    &\times\left[(J_j^\dag J_{j'}\rho-\rho J_{j}^\dag J_{j'}) \mathcal{P} \Big(\frac{1}{\omega_+ - \nu_k}\Big) -(J_j J_{j'}^\dag\rho -\rho J_{j}J_{j'}^\dag)\mathcal{P} \Big(\frac{1}{\omega_+ + \nu_k}\Big)\right]\\
    &+\frac{\omega_-^3 z_\mathrm{ho}^2 |\mathbf{d}|^2\pi}{2\epsilon_0\hbar(2\pi)^3 c ^3}\sum_{jj'}\left(J_{j'}\rho J_{j}^\dag-\frac{1}{2}\left\{J_j^\dag J_{j'},\rho\right\} \right)\int_0^{2\pi}\mathrm{d}\phi\int_0^\pi\mathrm{d}\theta \sin^3\theta(\mathbf{k_-}\cdot\hat{z})^2 e^{\mathrm{i}\frac{\omega_+}{c}\hat{k}\cdot\mathbf{r}^0_{jj'}}\\
    &-\frac{\mathrm{i}|\mathbf{d}|^2z_\mathrm{ho}^2}{4\epsilon_0\hbar(2\pi)^3 c^3}\sum_{jj'}\int_0^\infty \mathrm{d}\nu_k\nu_k^3\int_0^{2\pi}\mathrm{d}\phi\int_0^\pi\mathrm{d}\theta \sin^3\theta (\mathbf{k}\cdot\hat{z})^2e^{\mathrm{i}\mathbf{k}\cdot\mathbf{r}^0_{jj'}}\\
    &\times\left[(L_j^\dag L_{j'}\rho-\rho L_{j}^\dag L_{j'}) \mathcal{P} \Big(\frac{1}{\omega_- - \nu_k}\Big) -(L_j L_{j'}^\dag\rho -\rho L_{j}L_{j'}^\dag)\mathcal{P} \Big(\frac{1}{\omega_- + \nu_k}\Big)\right].
\end{split}
\end{equation*}
We need to perform next the integral over the angular variables in $\mathbf{k}$-space. For the particular situation we consider in this paper, the axis between the atoms is parallel to the direction of the displacement in the trap, i.e. $\mathbf{r}_{jj'}^0=r_{jj'}^0\hat{z}$. This allows us to write
\begin{equation*}
    \int_0^{2\pi}\mathrm{d}\phi\int_0^\pi\mathrm{d}\theta \sin^3\theta (\mathbf{k}\cdot\hat{z})^2 e^{\mathrm{i}\mathbf{k}\cdot\mathbf{r}^0_{jj'}}=\frac{1}{(r_{jj'}^0)^2}\int_0^{2\pi}\mathrm{d}\phi\int_0^\pi\mathrm{d}\theta \sin^3\theta (\mathbf{k}\cdot\mathbf{r}^0_{jj'})^2 e^{\mathrm{i}\mathbf{k}\cdot\mathbf{r}^0_{jj'}}.
\end{equation*}
which notably simplifies the calculation of this integral, since we can use the relation
\begin{equation*}
    (\mathbf{k}\cdot\mathbf{r}^0_{jj'})^2 e^{\mathrm{i}\mathbf{k}\cdot\mathbf{r}^0_{jj'}}=(kr_{jj'}^0)^2(\hat{k}\cdot\hat{r}^0_{jj'})^2 e^{\mathrm{i}(kr_{jj'}^0)^2(\hat{k}\cdot\hat{r}^0_{jj'})^2}=-(kr_{jj'}^0)^2\frac{\partial^2}{\partial (kr_{jj'}^0)^2}e^{\mathrm{i}(kr_{jj'}^0)^2(\hat{k}\cdot\hat{r}^0_{jj'})^2},
\end{equation*}
such that
\begin{equation*}
\begin{split}
    \int_0^{2\pi}\mathrm{d}\phi\int_0^\pi\mathrm{d}\theta \sin^3\theta (\mathbf{k}\cdot\hat{z})^2 e^{\mathrm{i}\mathbf{k}\cdot\mathbf{r}^0_{jj'}}=&-k^2\frac{\partial^2}{\partial (kr_{jj'}^0)^2}\int_0^{2\pi}\mathrm{d}\phi\int_0^\pi\mathrm{d}\theta \sin^3\theta e^{\mathrm{i}\mathbf{k}\cdot\mathbf{r}^0_{jj'}}\\=&-4\pi k^2\frac{\partial^2}{\partial (kr_{jj'}^0)^2}F(kr_{jj'}^0)\equiv-4\pi k^2 F''(kr_{jj'}^0).
\end{split}
\end{equation*}
Note, however, that this result is only valid for the case $\mathbf{r}_{jj'}^0=r_{jj'}^0\hat{z}$ where $\hat{z}$ is the direction of the atom vibration in the trap. In the more general case where the two vectors are not parallel, the calculation of this integral is more involved.

Introducing the result of the angular integral in our master equation we obtain
%and, considering that
%\begin{equation*}
%    \int_0^{2\pi}\mathrm{d}\phi\int_0^\pi\mathrm{d}\theta \sin^3\theta\cos^2\theta e^{\mathrm{i}\frac{\omega_0}{c}\hat{k}\cdot\mathbf{r}^0_{jj'}}=4\pi F''(k_0r^0_{jj'}),
%\end{equation*}
%with
%\begin{equation*}
%\begin{split}
%    F''(\kappa_{jj'})=-\partial^2_{\kappa_{jj'}} F(\kappa_{jj'})=&\left[1-(\hat{d}\cdot\hat{r}^0_{jj'})^2\right]\left[2\frac{\cos{\kappa_{jj'}}}{\kappa_{jj'}^2}+\left(\frac{1}{\kappa_{jj'}}-\frac{2}{\kappa_{jj'}^3}\right)\sin\kappa_{jj'}\right]\\&+\left[1-3(\hat{d}\cdot\hat{r}^0_{jj'})^2\right]\left[\left(\frac{1}{\kappa_{jj'}^2}-\frac{12}{\kappa_{jj'}^4}\right)\cos\kappa_{jj'}+\left(\frac{12}{\kappa_{jj'}^5}-\frac{5}{\kappa_{jj'}^3}\right)\sin\kappa_{jj'}\right],
    %\end{split}
%\end{equation*}
%where we have introduced $\kappa_{jj'}=k_0r^0_{jj'}$, we obtain
\begin{equation*}
    \begin{split}
    \dot{\rho}=&-\gamma\eta_0^2\sum_{jj'}\frac{3}{2}F''(k_0r^0_{jj'})\left(J_{j'}\rho J_{j}^\dag-\frac{1}{2}\left\{J_j^\dag J_{j'},\rho\right\}+L_{j'}\rho L_{j}^\dag-\frac{1}{2}\left\{L_j^\dag L_{j'},\rho\right\} \right)\\
    &+\frac{\mathrm{i}|\mathbf{d}|^2\eta_0^2}{3\epsilon_0\hbar c^5\pi^2k_0^2}\sum_{jj'}\frac{3}{4}\int_0^\infty \mathrm{d}\nu_k\nu_k^5 F''\left(\frac{\nu_k}{c}r^0_{jj'}\right)\left[(J_j^\dag J_{j'}\rho-\rho J_{j}^\dag J_{j'}) \mathcal{P} \Big(\frac{1}{\omega_0 - \nu_k}\Big) -(J_j J_{j'}^\dag\rho -\rho J_{j}J_{j'}^\dag)\mathcal{P} \Big(\frac{1}{\omega_0 + \nu_k}\Big)\right.\\
    &\left.(L_j^\dag L_{j'}\rho-\rho L_{j}^\dag L_{j'}) \mathcal{P} \Big(\frac{1}{\omega_0 - \nu_k}\Big) -(L_j L_{j'}^\dag\rho -\rho L_{j}L_{j'}^\dag)\mathcal{P} \Big(\frac{1}{\omega_0 + \nu_k}\Big)\right],
\end{split}
\end{equation*}
with $\eta_0=k_0 z_\mathrm{ho}/\sqrt{2}$, and where we have assumed that $\omega_+\approx\omega_-\approx\omega_0$. For the second term, we use that $F''$ is an even function of the frequency, such that
\begin{equation*}
    \mathcal{P}\int_0^\infty \mathrm{d}\nu_k\nu_k^5 F''\left(\frac{\nu_k}{c}r^0_{jj'}\right)\left[ \frac{1}{\nu_k+\omega_0} +\frac{1}{ \nu_k-\omega_0}\right]=\mathcal{P}\int^\infty_{-\infty} \mathrm{d}\nu_k F''\left(\frac{\nu_k}{c}r^0_{jj'}\right) \frac{\nu_k^5}{\nu_k-\omega_0}=\pi\omega_0^5G''\left(k_0r^0_{jj'}\right),
\end{equation*}
with
\begin{align*}
    G''(kr_{jj'}^0)=\frac{\partial^2}{\partial (kr_{jj'}^0)^2}G(kr_{jj'}^0)%=&\left[1-(\hat{d}\cdot\hat{r}^0_{jj'})^2\right]\left[\cos\kappa_{jj'}\left(\frac{1}{\kappa_{jj'}}-\frac{2}{\kappa_{jj'}^2}\right)\right]\\
    %&-\left[1-3(\hat{d}\cdot\hat{r}^0_{jj'})^2\right]\left[\left(\frac{1}{\kappa_{jj'}^2}-\frac{12}{\kappa_{jj'}^4}\right)\sin\kappa_{jj'}+\left(\frac{12}{\kappa_{jj'}^5}-\frac{5}{\kappa_{jj'}^3}\right)\cos\kappa_{jj'}\right].
\end{align*}
Finally, making the definitions $\Gamma''_{jj'}=\frac{3\gamma}{2} F''(k_0r^0_{jj'})$ and $V''_{jj'}=\frac{3\gamma}{4} G''(k_0r^0_{jj'})$, we obtain
\begin{equation*}
    \begin{split}
    \dot{\rho}=&-\eta_0^2\sum_{jj'}\Gamma''_{jj'}\left(J_{j'}\rho J_{j}^\dag-\frac{1}{2}\left\{J_j^\dag J_{j'},\rho\right\} \right)-\mathrm{i}\eta_0^2\sum_{j\neq j'}V''_{jj'}\left[J_{j'}^\dag J_j,\rho\right]\\
    &-\eta_0^2\sum_{jj'}\Gamma''_{jj'}\left(L_{j'}\rho L_{j}^\dag-\frac{1}{2}\left\{L_j^\dag L_{j'},\rho\right\} \right)-\mathrm{i}\eta_0^2\sum_{j\neq j'}V''_{jj'}\left[L_{j'}^\dag L_j,\rho\right],
    \end{split}
\end{equation*}
where we have again considered only the terms $V''_{jj'}$ with $j\neq j'$, absorbing the $j=j'$ energy shift into the atomic energy.

\subsection{Terms proportional to $\eta^2$: $\left[H_0(t),\left[H_2^0(t-\tau),\rho\right]\right]$ and $\left[H_2^0(t),\left[H_0(t-\tau),\rho\right]\right]$}

Starting again from the Redfield equation, and introducing the notation
\begin{equation*}
    C_j=1+2a_j^\dag a_j,\qquad
    \tilde{A}_j(t)=A_j(t)C_j
\end{equation*}
we have
\begin{equation*}
    \begin{split}
    \dot{\rho}=&-\frac{1}{\hbar^2}\int_0^\infty \mathrm{d}\tau \mathrm{Tr}_E\left\{\left[H_0(t),\left[H_2^0(t-\tau),\rho(t)\otimes\rho_E\right]\right]\right\}\\=&\!\!\!\!\!\!\sum_{\mathbf{k}\mathbf{k}'jj'\lambda\lambda'}\frac{\eta^2}{2}g^\lambda_\mathbf{k}{g^{\lambda'}_{\mathbf{k}'}}^*\int_0^\infty\mathrm{d}\tau\mathrm{Tr}_E\left\{\left[A_j(t)\left(B_{\lambda\mathbf{k}j}(t)+B^\dag_{\lambda\mathbf{k}j}(t)\right),\left[\tilde{A}_{j'}(t-\tau)\left(B_{\lambda'\mathbf{k}'j'}(t-\tau)+B^\dag_{\lambda'\mathbf{k}'j'}(t-\tau)\right),\rho(t)\otimes\rho_E\right]\right]\right\}\\
    =&\!\!\!\!\!\!\sum_{\mathbf{k}\mathbf{k}'jj'\lambda\lambda'}\frac{\eta^2}{2}g^\lambda_\mathbf{k}{g^{\lambda'}_{\mathbf{k}'}}^*\int_0^\infty \mathrm{d}\tau\mathrm{Tr}_E\left\{\left[A_j(t)\left(B_{\lambda\mathbf{k}j}(t)+B^\dag_{\lambda\mathbf{k}j}(t)\right),\tilde{A}_{j'}(t-\tau)\left(B_{\lambda'\mathbf{k}'j'}(t-\tau)+B^\dag_{\lambda'\mathbf{k}'j'}(t-\tau)\right)\rho(t)\otimes\rho_E\right]\right.\\
    &\left.-\left[A_j(t)\left(B_{\lambda\mathbf{k}j}(t)+B^\dag_{\lambda\mathbf{k}j}(t)\right),\rho(t)\otimes\rho_E\tilde{A}_{j'}(t-\tau)\left(B_{\lambda'\mathbf{k}'j'}(t-\tau)+B^\dag_{\lambda'\mathbf{k}'j'}(t-\tau)\right)\right]\right\}\\=&\!\!\!\!\!\!\sum_{\mathbf{k}\mathbf{k}'jj'\lambda\lambda'}\frac{\eta^2}{2} g^\lambda_\mathbf{k}{g^{\lambda'}_{\mathbf{k}'}}^*\int_0^\infty \mathrm{d}\tau\mathrm{Tr}_E\left\{A_j(t)\tilde{A}_{j'}(t-\tau)\left(B_{\lambda\mathbf{k}j}(t)+B^\dag_{\lambda\mathbf{k}j}(t)\right)\left(B_{\lambda'\mathbf{k}'j'}(t-\tau)+B^\dag_{\lambda'\mathbf{k}'j'}(t-\tau)\right)\rho(t)\otimes\rho_E\right.\\
    &-\tilde{A}_{j'}(t-\tau)\left(B_{\lambda'\mathbf{k}'j'}(t-\tau)+B^\dag_{\lambda'\mathbf{k}'j'}(t-\tau)\right)\rho(t)\otimes\rho_EA_j(t)\left(B_{\lambda\mathbf{k}j}(t)+B^\dag_{\lambda\mathbf{k}j}(t)\right)\\
    &-A_j(t)\left(B_{\lambda\mathbf{k}j}(t)+B^\dag_{\lambda\mathbf{k}j}(t)\right)\rho(t)\otimes\rho_E\tilde{A}_{j'}(t-\tau)\left(B_{\lambda'\mathbf{k}'j'}(t-\tau)+B^\dag_{\lambda'\mathbf{k}'j'}(t-\tau)\right)\\
    &\left.+\rho(t)\otimes\rho_E\tilde{A}_{j'}(t-\tau)\left(B_{\lambda'\mathbf{k}'j'}(t-\tau)+B^\dag_{\lambda'\mathbf{k}'j'}(t-\tau)\right)A_j(t)\left(B_{\lambda\mathbf{k}j}(t)+B^\dag_{\lambda\mathbf{k}j}(t)\right)\right\}.
    \end{split}
\end{equation*}
Taking again the expectation value with respect to the environment, i.e. the radiation field at zero temperature, we obtain
\begin{equation*}
    \begin{split}
    \dot{\rho}=&\sum_{\mathbf{k}\lambda jj'}\frac{\eta^2}{2}|g^\lambda_\mathbf{k}|^2\int_0^\infty \mathrm{d}\tau\left\{\left[A_j(t)\tilde{A}_{j'}(t-\tau)\rho(t)-\tilde{A}_{j'}(t-\tau)\rho(t)A_j(t)\right]e^{\mathrm{i}\left(\mathbf{k}\cdot\mathbf{r}^0_{jj'}-\nu_k\tau\right)}+\right.\\
    &\left.\left[\rho(t)\tilde{A}_{j'}(t-\tau)A_j(t)-A_j(t)\rho(t)\tilde{A}_{j'}(t-\tau)\right]e^{-\mathrm{i}\left(\mathbf{k}\cdot\mathbf{r}^0_{jj'}-\nu_k\tau\right)}\right\}.
    \end{split}
\end{equation*}
We write out the $\tilde{A}_j(t)$ terms and perform the secular approximation, neglecting all terms that contain a exponential function that depends on the time $t$, for example,
\begin{align*}
    A_j(t)\tilde{A}_{j'}(t-\tau)\rho(t)=&(e^{i\omega_0 t}\sigma_j^\dag+e^{-i\omega_0 t}\sigma_j)(e^{i\omega_0 t}e^{-i\omega_0 \tau}\sigma_{j'}^\dag+e^{-i\omega_0 t}e^{i\omega_0 \tau}\sigma_{j'})(1+2a_{j'}^\dag a_{j'})\rho\\
    \approx& (e^{-i\omega_0 \tau}\sigma_j\sigma_{j'}^\dag+e^{i\omega_0 \tau}\sigma_j^\dag \sigma_{j'})(1+2a_{j'}^\dag a_{j'})\rho.
\end{align*}
Now we can write, using the shortcut notation $K_j=\sigma_j (1+2a^\dag_j a_j)$,
\begin{equation*}
    \begin{split}
    \dot{\rho}=&\sum_{\mathbf{k}\lambda jj'}\frac{\eta^2}{2}|g^\lambda_\mathbf{k}|^2\int_0^\infty \mathrm{d}\tau\left\{\left[(\sigma_j^\dag K_{j'}\rho-K_{j'}\rho\sigma_j^\dag) e^{i(\omega_0-\nu_k) \tau}+(\sigma_j K_{j'}^\dag\rho -K_{j'}^\dag\rho\sigma_j)e^{-i(\omega_0+\nu_k)\tau}\right]e^{\mathrm{i}\mathbf{k}\cdot\mathbf{r}^0_{jj'}}+\right.\\
    &\left.\left[(\rho K_{j'}\sigma_j^\dag-\sigma_j^\dag \rho K_{j'})e^{i(\omega_0+\nu_k)\tau}+(\rho K_{j'}^\dag\sigma_j -\sigma_j\rho K_{j'}^\dag) e^{-i(\omega_0-\nu_k)\tau}\right]e^{-\mathrm{i}\mathbf{k}\cdot\mathbf{r}^0_{jj'}}\right\}\\
    =&\sum_{\mathbf{k}\lambda jj'}\frac{\eta^2}{2}|g^\lambda_\mathbf{k}|^2e^{\mathrm{i}\mathbf{k}\cdot\mathbf{r}^0_{jj'}}\int_0^\infty \mathrm{d}\tau\left[(\sigma_j^\dag K_{j'}\rho-K_{j'}\rho\sigma_j^\dag) e^{i(\omega_0-\nu_k) \tau}+(\sigma_j K_{j'}^\dag\rho -K_{j'}^\dag\rho\sigma_j)e^{-i(\omega_0+\nu_k)\tau}\right.\\
    &\left.+(\rho K_{j}\sigma_{j'}^\dag-\sigma_{j'}^\dag \rho K_{j})e^{i(\omega_0+\nu_k)\tau}+(\rho K_{j}^\dag\sigma_{j'} -\sigma_{j'}\rho K_{j}^\dag) e^{-i(\omega_0-\nu_k)\tau}\right],
   \end{split}
\end{equation*}
where in the last step we have exchanged $j\to j'$ in the second lines. All of the mathematical steps that follow this point are the same as the ones used in the previous two sections and hence are only sketched. We use first the Heitler function and obtain
\begin{equation*}
    \begin{split}
    \dot{\rho}=&\sum_{\mathbf{k}\lambda jj'}\frac{\eta^2}{2}|g^\lambda_\mathbf{k}|^2e^{\mathrm{i}\mathbf{k}\cdot\mathbf{r}^0_{jj'}}\left[\pi \delta(\omega_0 - \nu_k)\left(\sigma_j^\dag K_{j'}\rho+\rho K^\dag_{j}\sigma_{j'} -K_{j'}\rho\sigma_{j}^\dag-\sigma_{j'}\rho K_{j}^\dag\right)\right.\\
    &\left.+\mathrm{i}\left[(\sigma_j^\dag K_{j'}\rho-\rho K^\dag_{j}\sigma_{j'}-K_{j'}\rho\sigma_j^\dag+\sigma_{j'}\rho K_j^\dag) \mathcal{P} \Big(\frac{1}{\omega_0 - \nu_k}\Big) -(\sigma_j K_{j'}^\dag\rho -\rho K_{j}\sigma_{j'}^\dag-K_{j'}^\dag\rho\sigma_j+\sigma_{j'}^\dag\rho K_j)\mathcal{P} \Big(\frac{1}{\omega_0 + \nu_k}\Big)\right]\right].
   \end{split}
\end{equation*}

In order to write this term of the master equation, we now proceed the same way with the terms containing $\left[H_2^0(t),\left[H_0(t-\tau),\rho\right]\right]$ to this expression. For symmetry reasons, these terms look identical to the ones we have already obtained but with the $K$ and $\sigma$ operators exchanged. Hence, adding these two terms we obtain the contribution
\begin{equation*}
    \begin{split}
    \dot{\rho}=&\sum_{\mathbf{k}\lambda jj'}\frac{\eta^2}{2}|g^\lambda_\mathbf{k}|^2e^{\mathrm{i}\mathbf{k}\cdot\mathbf{r}^0_{jj'}}\left\{\pi \delta(\omega_0 - \nu_k)\left(\sigma_j^\dag K_{j'}\rho+\rho K^\dag_{j}\sigma_{j'} -K_{j'}\rho\sigma_{j}^\dag-\sigma_{j'}\rho K_{j}^\dag+K_j^\dag \sigma_{j'}\rho+\rho \sigma^\dag_{j}K_{j'} -\sigma_{j'}\rho K_{j}^\dag-K_{j'}\rho \sigma_{j}^\dag\right)\right.\\
    &+\mathrm{i}\left[(\sigma_j^\dag K_{j'}\rho-\rho K^\dag_{j}\sigma_{j'}-K_{j'}\rho\sigma_j^\dag+\sigma_{j'}\rho K_j^\dag+K_j^\dag \sigma_{j'}\rho-\rho \sigma^\dag_{j}K_{j'}-\sigma_{j'}\rho K_j^\dag+K_{j'}\rho \sigma_j^\dag) \mathcal{P} \Big(\frac{1}{\omega_0 - \nu_k}\Big)\right. \\
    &\left.\left.-(\sigma_j K_{j'}^\dag\rho -\rho K_{j}\sigma_{j'}^\dag-K_{j'}^\dag\rho\sigma_j+\sigma_{j'}^\dag\rho K_j+K_j \sigma_{j'}^\dag\rho -\rho \sigma_{j}K_{j'}^\dag-\sigma_{j'}^\dag\rho K_j+K_{j'}^\dag\rho \sigma_j)\mathcal{P} \Big(\frac{1}{\omega_0 + \nu_k}\Big)\right]\right\} \\
    =&\sum_{\mathbf{k}\lambda jj'}\frac{\eta^2}{2}|g^\lambda_\mathbf{k}|^2e^{\mathrm{i}\mathbf{k}\cdot\mathbf{r}^0_{jj'}}\left\{\pi \delta(\omega_0 - \nu_k)\left(\sigma_j^\dag K_{j'}\rho+\rho \sigma^\dag_{j}K_{j'}-2K_{j'}\rho\sigma_{j}^\dag+K_j^\dag \sigma_{j'}\rho+\rho K^\dag_{j}\sigma_{j'} -2\sigma_{j'}\rho K_{j}^\dag\right)\right.\\
    &\left.+\mathrm{i}\left[(\sigma_j^\dag K_{j'}\rho-\rho K^\dag_{j}\sigma_{j'}+K_j^\dag \sigma_{j'}\rho-\rho \sigma^\dag_{j}K_{j'}) \mathcal{P} \Big(\frac{1}{\omega_0 - \nu_k}\Big)-(\sigma_j K_{j'}^\dag\rho -\rho K_{j}\sigma_{j'}^\dag+K_j \sigma_{j'}^\dag\rho -\rho \sigma_{j}K_{j'}^\dag)\mathcal{P} \Big(\frac{1}{\omega_0 + \nu_k}\Big)\right]\right\}.
   \end{split}
\end{equation*}
Absorbing again the $j=j'$ energy shift into the atomic energy, one obtains
\begin{equation*}
    \begin{split}
\dot{\rho}=&-\sum_{\mathbf{k}\lambda jj'}\frac{\eta^2}{2}|g^\lambda_\mathbf{k}|^2e^{\mathrm{i}\mathbf{k}\cdot\mathbf{r}^0_{jj'}}2\pi \delta(\omega_0 - \nu_k)\left(K_{j'}\rho\sigma_{j}^\dag-\frac{1}{2}\left\{\sigma_j^\dag K_{j'},\rho\right\}+\sigma_{j'}\rho K_{j}^\dag-\frac{1}{2}\left\{K_j^\dag \sigma_{j'},\rho\right\}\right)\\
    &-\mathrm{i}\sum_{\mathbf{k}\lambda j\neq j'}\frac{\eta^2}{2}|g^\lambda_\mathbf{k}|^2e^{\mathrm{i}\mathbf{k}\cdot\mathbf{r}^0_{jj'}}\left(\left[\sigma_j^\dag K_{j'},\rho\right]+\left[K_j^\dag \sigma_{j'},\rho\right]\right) \mathcal{P} \Big(\frac{1}{\nu_k - \omega_0}+\frac{1}{\omega_0 + \nu_k}\Big).
   \end{split}
\end{equation*}
Finally, we can write the final contribution to the dynamics of order $\eta_0^2$ as
\begin{equation*}
    \begin{split}
\dot{\rho}=&\frac{\eta_0^2}{2}\sum_{jj'}\Gamma''_{jj'}\left(K_{j'}\rho\sigma_{j}^\dag-\frac{1}{2}\left\{\sigma_j^\dag K_{j'},\rho\right\}+\sigma_{j'}\rho K_{j}^\dag-\frac{1}{2}\left\{K_j^\dag \sigma_{j'},\rho\right\}\right)\\
    &\mathrm{i}\frac{\eta_0^2}{2}\sum_{j\neq j'}V''_{jj'}\left(\left[\sigma_j^\dag K_{j'},\rho\right]+\left[K_j^\dag \sigma_{j'},\rho\right]\right).
   \end{split}
\end{equation*}
All of these contributions added together give rise to the final Lindblad equation that we provide in the main text.

%\bibliography{sample}